\documentclass[12pt]{iopart}
  
\usepackage{iopams}
\usepackage{cite}
\usepackage{graphicx}
\usepackage{amssymb}
\usepackage[british]{babel}
\usepackage{color}
\newcommand\eqref[1]{(\ref{#1})}
\usepackage[colorlinks=true,allcolors=blue]{hyperref}
\usepackage[T1]{fontenc}

\begin{document}

\title[Heterogeneous active particle dynamics]{Non-Gaussian displacement
distributions in models of heterogeneous active particle dynamics}

\author{Elisabeth Lemaitre$^{\dagger}$, Igor M. Sokolov$^{\flat}$, Ralf
Metzler$^{\dagger,\sharp}$, and Aleksei V. Chechkin$^{\dagger,\ddagger,\S}$}
\address{$\dagger$ Institute of Physics \& Astronomy, University of Potsdam,
14476 Potsdam-Golm, Germany\\
$\ddagger$Faculty of Pure and Applied Mathematics, Hugo Steinhaus Center,
Wroc{\l}aw University of Science and Technology, Wyspianskiego 27,
50-370 Wroc{\l}aw, Poland\\
$\S$Akhiezer Institute for Theoretical Physics, 61108 Kharkov, Ukraine\\
$\flat$ Institute of Physics, Humboldt University Berlin, Newtonstrasse 15,
D-12489 Berlin, Germany\\
and IRIS Adlershof, Zum Gro{\ss}en Windkanal 6, 12489 Berlin, Germany\\
$\sharp$ Asia Pacific Centre for Theoretical Physics, Pohang 37673,
Republic of Korea}

\begin{abstract}
We study the effect of randomly distributed diffusivities and speeds in two
models for active particle dynamics with active and passive fluctuations. We
demonstrate how non-Gaussian displacement distributions emerge in these models
in the long time limit, including Cauchy-type and exponential (Laplace)
shapes. Notably the resulting shapes of the displacement distributions
with distributed diffusivities for the active models considered here are
in striking contrast to passive diffusion models. For the active motion
models our discussion points out the differences between active- and
passive-noise. Specifically, we demonstrate that the case with active-noise
is in nice agreement with measured data for the displacement distribution
of social amoeba.
\end{abstract}

\section{Introduction}

Brownian motion \cite{brown}, the thermally driven pedesis of colloidal
particles in liquid environments, was a key phenomenon in developing
statistical physics \cite{landau,schwabl}. Specifically, Einstein
and Smoluchowski \cite{einstein,smoluchowski} realised that while the
instantaneous velocity of colloidal particles changes too rapidly and
thus escapes experimental measurement, the decisive observable quantity
is the mean displacement,\footnote{Einstein's writes: "Die mittlere
Verschiebung ist also proportional der Quadratwurzel der Zeit", in the
sense of the standard deviation \cite{einstein}.} for which they obtained
the characteristic square-root time dependence. Following the time scale
separation of the velocity dynamics and the observed particle displacement, the
diffusion coefficient was shown to be proportional to thermal energy $k_BT$
\cite{einstein,smoluchowski}. This fact was employed by experimentalists
like Perrin and Nordlund to deduce Avogadro's number from single particle
tracking measurements of Brownian particles \cite{perrin,nordlund}, while
Kappler used torsional Brownian motion to map out the associated Gaussian
displacement distribution \cite{kappler}.

Augmenting Newton's second law with a fluctuating force, now a central
concept of non-equilibrium statistical physics \cite{landau1,brenig},
Langevin \cite{langevin} explicitly included inertia and damping effects.
At short times, such a particle moves ballistically, while after many
random changes of the particle direction, at longer times the mean
squared displacement (MSD) assumes the "diffusive" linear scaling in time.
Neglecting the mass term in the Langevin equations leads to the "overdamped
limit" \cite{zwanzig,vankampen}. The deterministic description for the
probability density function (PDF) of the Brownian particle is provided by
the diffusion equation \cite{vankampen}. The crossover from ballistic to
diffusive behaviour of the MSD encoded in the full Langevin equation is
captured by the Klein-Kramers equation for the bivariate PDF of velocity
and position \cite{klein,kramers,risken}.

In a non-equilibrium setting, by "active motion" self-propelling particles are
able to maintain a finite, directed velocity over experimentally observable
time scales. Examples for such active particles include microswimmers such as
bacteria \cite{microswim,detailed_balance,bacteria}, amoeba \cite{amoeba},
sperm cells \cite{gert}, or flagellates \cite{stark,protozoa}, see
\cite{Romanczuk,marchetti} for reviews on active motion.
Self-propulsion can also be realised in colloidal "Janus particles"
with non-isotropic surfaces activated \cite{ramin}, for instance, through
light-induced diffusio-osmotic flows \cite{santer1,santer2} or diffusiophoresis
\cite{lowen}. Active motion is crucial for the proper functioning of many
biological processes such as intracellular transport via molecular motors
\cite{seisenhuber,elbaum,granick1}. In the absence of an external bias such as
drift or chemotactic fields there are finite-time correlations in the motion
of such active particles. Consequently, after times significantly longer than
this correlation time the direction of the particle becomes randomised, and
the displacement PDF will approach a Gaussian \cite{Romanczuk,microswim,
detailed_balance,bacteria}. The resulting diffusive dynamics is
characterised by a linear-in-time MSD with an effective diffusion coefficient.

Deviations from the linear time dependence of the MSD and the Gaussian
displacement PDF, observed for Brownian motion and for active particles
beyond the correlation time of the initially persistent motion, occur quite
widely, in a large variety of systems. One deviation, "anomalous diffusion",
typically refers to power-law time dependencies of the MSD, which may be
caused, inter alia, by spatial and energetic disorder or viscoelastic effects
\cite{bouchaud,report,igor,pccp}.

The other deviation is the occurrence of non-Gaussian displacement
distributions, despite a linear time dependence (Fickian behaviour) of
the MSD \cite{granick,erice}. This phenomenon is observed, inter alia,
for the motion of submicron tracers along linear tubes and in entangled
actin networks \cite{granick2}, tracer dynamics in hard sphere colloidal
suspensions \cite{granick3}, diffusion of nanoparticles in nanopost or
micropillar arrays \cite{post,yael}, and for tracers in mucin hydrogels
in certain parameter ranges \cite{mucin,mucin1}. Prominent physical models
explaining Fickian yet non-Gaussian diffusion are based on the assumption
of a heterogeneous environment. In a mean-field, annealed sense the value
of the particle diffusivity in different patches of the environment is
captured by a "superstatistical", time-independent PDF \cite{beck,granick},
similar to "grey Brownian motion" models \cite{grey,grey1}. Alternatively,
a stochastically varying instantaneous particle diffusivity is assumed in
"diffusing diffusivity" models \cite{gary,kls,tyagi,diffdiff,vit,vit1,yann,yann1}.
The latter are characterised by a finite correlation time, beyond which a
long-time Gaussian displacement PDF with effective diffusion coefficient
emerges. We note that time-varying instantaneous diffusivities naturally
occur in perpetually (de)polymerising molecules \cite{mario,fulvio,fulvio1}
or in shape-shifting proteins \cite{eiji}. We also mention that Brownian
yet non-Gaussian diffusion emerges from random-coefficient autoregressive
models \cite{chris}. The quenched nature of the
environment, relevant when the environment is static or changing slowly on
the time scale of the particle motion \cite{jakubstas}, can be considered
in numerical approaches \cite{liu,liu1,zhenya,pacheco}. Finally, extreme value
arguments can be employed to explain the emergence of finite-time
exponential PDFs \cite{eli,eli1}. Non-Gaussian statistics in combination
with viscoelastic diffusion, a generically Gaussian process, have been
observed, e.g., in cellular cytoplasm \cite{bac_cyto}, in mucin gels
\cite{mucin,mucin1} for certain parameter ranges, the motion of lipids in
protein-crowded bilayer membranes \cite{ilpo}, as well as in crowded media
\cite{surya,kolja}, for drug molecules diffusing between silica nanoslabs
\cite{amanda}, or for Lennard-Jones particles \cite{sandalo}. We finally
mention non-Gaussian tracer diffusion in active
gels \cite{active_gels}. For non-Gaussian, long-range correlated motion such
as viscoelastic diffusion, superstatistical approaches \cite{jakub} as
well as diffusing diffusivity models \cite{wei} have been proposed.

Here we are interested in \emph{non-Gaussian displacement statistics of
actively moving particles}. Experimental observations of such non-Gaussianity
include polymers in active particle baths \cite{jaeoh},
microswimmers \cite{activebath2}, social amoeba
\cite{beta}, self-propelling Janus particles \cite{lowen}, progenitor
cells \cite{runtumble}, or nematodes \cite{hapca}. As mentioned above,
the typically employed models for active motion predict a crossover from
short time directed motion to a Brownian dynamics with Gaussian displacement
PDF and effective diffusion coefficient, beyond the correlation time for the
persistence of the motion \cite{microswim,detailed_balance,bacteria,Romanczuk}.
The occurrence of long-time, non-Gaussian statistics may either be due to
variations of the individual speeds or the persistence of the organisms,
or to heterogeneities in the environment. To account for non-Gaussianity on
the level of independently moving active particles, in what follows we
complement a recent study of active motion in a
heterogeneous environment based on a randomly evolving viscosity \cite{khadem},
and here introduce two possible extensions of distributed parameters: (i)
random speeds, reflecting natural variations from one organism to the next,
and (ii) randomly distributed effective diffusivities of the particles
reflecting heterogeneous environments or different shapes and sizes of the
particles. We will analyse the emerging motion caused by these
randomised parameters, in two different active particle models: (i)
active particles with passive-noise (APPs) and (ii) active Brownian particles
(ABPs) with active-noise. We will study the resulting displacement PDFs,
including exponential, stretched Gaussian, and Cauchy-type shapes. Comparison
with experiments of social amoeba shows nice agreement with our random-speed
ABP model.

After introducing the APP and ABP models in section \ref{sec2}, we then
formulate and analyse superstatistical generalisations of these models
in section \ref{sec3}. We summarise our results in section \ref{sec4} and
present details on derivations in the Appendices.

\section{Model}
\label{sec2}

Active matter describes a broad range of systems ranging, inter alia,
from molecular motors \cite{molecmotor} over colloid-size bacteria
\cite{bacteria} to larger organisms or particles such as insects or fish
\cite{klamser} (or more generally agent-based prey-predator interactions
\cite{klamser2}). These systems are all characterised by self-propulsion, and
they undergo fluctuations either due to external \cite{khadem} or internal
\cite{peruani} processes. Several models were developed to represent these
organisms with various ranges of applicability. We will consider in the
following active motion with active and passive fluctuations, and take the size
of the system to be sufficiently sparse such that we can ignore collective
effects. In this context, we refer to passive fluctuations in the sense
defined in \cite{PRL_Rom}: passive-noise is independent of the orientation
of the particle, whereas active-noise acts parallel or perpendicular to the
orientation of the particle.  As we will see this distinction has significant
consequences in the respective particle dynamics.  We will restrict the
discussion to the case of constant propulsion speed for a given individual,
which is a valid assumption as long as the speed remains much larger than
the fluctuations \cite{Romanczuk}.

\subsection{Active particles with passive-noise (APP)}

We consider an active agent moving in two dimensions at a speed $v(t)$ in
the direction $\mathbf{e_v}=(\cos\phi,\sin\phi)$. The present model starts
from the Langevin equation for the angle $\phi(t)$, obtained through
projection on $\mathbf{e_{\phi}}$ ($\mathbf{e_v}\cdot\mathbf{e_{\phi}}=0$)
of the underdamped Langevin equation of an active massive particle without
external forcing,
\begin{equation}
\label{vdyn}
m\dot{\mathbf{v}}(t)=-m\gamma\mathbf{v}(t)+\sqrt{2D}m\boldsymbol{\xi}(t).
\end{equation}
Here $\gamma$ is the friction coefficient of the particle of dimension $1/
\mathrm{time}$. While $\gamma$ is a constant for a passive particle, it
should be considered
velocity dependent in order to model self-propulsion effects \cite{Romanczuk}.
$\boldsymbol{\xi}(t)$ is a two-dimensional, zero-mean, Gaussian white noise
process with $\langle\xi_i(t_1)\xi_j(t_2)\rangle=\delta_{ij}\delta(t_1-t_2)$,
where $i$ and $j$ are the Cartesian coordinates $x$ and $y$. Finally, $D$
quantifies the noise strength. In the case of thermal noise one would get
$D=k_BT\gamma/m=\gamma^2D_T$, where $D_T$ denotes the classical Brownian
translational diffusivity \cite{complex_crowded}. For the remainder of this
work we assume the constant speed limit $\mathbf{v}(t)=v_0\mathbf{e_v}(t)$,
for which equation (\ref{vdyn}) for the velocity dynamics reduces to the
stochastic equation
\begin{equation}
\frac{d\phi(t)}{dt}=\frac{\sqrt{2D}}{v_0}\Big(\xi_y(t)\cos\phi(t)-
\xi_x(t)\sin\phi(t)\Big)
\label{phi_eq}
\end{equation}
for the director angle $\phi(t)$. While the rotational noise
$\xi_{\phi} (t)=(\xi_y(t)\cos\phi(t)-\xi_x(t)\sin \phi(t))$ itself
remains equivalent to a one-dimensional white noise process with
zero-mean $\langle\xi_{\phi}(t)\rangle=0$ and $\delta$-correlation
$\langle\xi_{\phi}(t_1)\xi_{\phi}(t_2)\rangle=\delta(t_1- t_2)$, due to the
coupling of the angular motion to the propulsion speed the stochastic equation
(\ref{phi_eq}) for the angle $\phi(t)$ includes the factor $1/v_0$. The
effective fluctuations considered in relation (\ref{phi_eq}) thus result
from the projection of the noise onto $\mathbf{e_v}$ and $\mathbf{e_{\phi}}$
and are called passive fluctuations, in contrast to active fluctuations, that
are understood to directly act in or perpendicular to the direction of motion,
$\mathbf{e_v}=\mathbf{v}/|\mathbf{v}|$ \cite{PRL_Rom}.

The APP model introduced here was originally studied by Mikhailov and
Meink{\"o}hn
\cite{Mikhailov1997SelfmotionIP} and yields the MSD (see \ref{letphi} and
\ref{appc} for the derivation)\footnote{We assume that the initial value
of the angle $\phi$ is uniformly distributed on $(0,2\pi)$ such that the
first moment is identically zero.}
\begin{equation}
\langle\mathbf{r}^2(t)\rangle=\frac{2v_0^4t}{D}+\frac{2v_0^6 }{D^2}\left[
\exp\left(-\frac{Dt}{v_0^2}\right)-1\right],
\label{MSD_U}
\end{equation}
which exhibits the ballistic scaling $\langle\Delta\mathbf{r}^2\rangle\sim v_0
^2t^2$ at short times and in the long time limit converges to the normal-diffusive
behaviour\footnote{We note that equation \eqref{MSD_U} does not exactly match
the MSD reported in \cite{Mikhailov1997SelfmotionIP}, which was found to
include two small typos, and our diffusivity $D$ is equivalent to their
expression $\sigma/m^2$.}
\begin{equation}
\langle\mathbf{r}^2\rangle\sim\frac{2v_0^4t}{D}\equiv4D_{\mathrm{eff}}t.
\end{equation}
Here $\tau=v_0^2/D$ is the typical
"persistence time" scale of the directed motion, corresponding to the
typical time over which directional memory is lost. In the diffusive
regime (i.e., on time scales exceeding the persistence time $\tau$) one
retrieves a Gaussian PDF with the effective diffusion coefficient
\cite{igorlutz}
\begin{equation}
\label{deff}
D_{\mathrm{eff}}=\frac{v_0^4}{2D}.
\end{equation}
Note the characteristic quartic dependence of $D_{\mathrm{eff}}$ on the speed
$v_0$ in this APP model. Typically the APP model is used for the description
of larger agents
with clear inertial effects, e.g., fish or birds, however, in low-viscosity
environments such as gases its relevance also extends to small-sized particles
\cite{Romanczuk}.

\subsection{Active Brownian particles (ABP)}

The alternative representation involving active fluctuations, the "minimal
ABP model", is more commonly found in literature \cite{ABP1,ABP2}. In this
description, uncorrelated white noise acts in parallel or perpendicular to
the time dependent orientation of the particle. The Langevin equations
describing the model are
\begin{eqnarray}
\label{ActiveOverdamped1}
\frac{d\mathbf{r}(t)}{dt}&=&v_0\mathbf{e_v}+\sqrt{2D_T}\boldsymbol{\xi}_T(t),\\
\frac{d\phi(t)}{dt}&=&\sqrt{2D_R}\xi_R(t).
\label{ActiveOverdamped}
\end{eqnarray}
Here $\boldsymbol{\xi}_T$ and $\xi_R$ are independent, uncorrelated, zero-mean
Gaussian white noise processes for translation and rotation, with $\langle\xi_{
T,i}(t_1)\xi_{T,j}(t_2)\rangle=\delta_{i,j}\delta(t_1-t_2)$ and $\langle\xi_R(t_1')\xi_R(t_2')\rangle=\delta(
t_1'-t_2')$.\footnote{For spherical colloids immersed in a thermal bath,
$D_T=k_BT/
(m\gamma)$, where $\gamma=6\pi\eta R/m$ in terms of the dynamic viscosity $\eta$
and the particle radius $R$, the Stokes-Einstein relation is fulfilled when
$D_R=3D_T/(4R^2)$ \cite{complex_crowded}.} The ABP model has been extensively
studied and the short and long-time MSD reads \cite{sevilla}
\begin{equation}
\langle\mathbf{r}^2(t)\rangle\sim\left\{\begin{array}{ll}\displaystyle
4D_Tt+(2D_TD_R+v_0^2)t^2,&\mbox{ for }t\ll1/D_R,\\[0.32cm]
\displaystyle 4\left(D_T+\frac{v_0^2}{2D_R}\right)t,&\mbox{ for }t\gg1/D_R.
\end{array}\right.
\end{equation}
According to this result the particles undergo up to three dynamic regimes.
Namely, in the short time regime $t\ll1/D_R$ we observe a crossover from an
initial linear scaling in time to a ballistic scaling as long as $4D_T/(2D_T
D_R+v_0^2)\leq1/D_R$ (i.e., $2D_TD_R\leq v_0^2$). This crossover then occurs
when $t\approx4D_T/(2D_TD_R+v_0^2)$. Otherwise the separate ballistic regime
does not occur. In the long time limit $t\gg 1/D_R$ a diffusive regime with
effective diffusivity
\begin{equation}
\label{deffprime}
D'_{\mathrm{eff}}=D_T+ v_0^2/(2D_R)
\end{equation}
is reached. The correlations time scale for the directed motion in the ABP
model is therefore identified as $\tau'=1/D_R$.

Despite their apparent similarity, these two active motion models actually
exhibit significant differences, notably the short-time diffusive behaviour
in the latter, arising from the translational diffusion, and the respective
dependency of the long-time effective diffusivity on the propulsion speed,
$D'_{\mathrm{eff}}\propto v_0^2$ (when the translational diffusivity can be
neglected) in the ABP model, in contrast to the quartic $v_0$-dependence of
$D_{\mathrm{eff}}$ in the APP case. No translational diffusivity acts in the
APP model, while the amplitude of the rotational diffusivity is inversely
proportional to the speed of the diffusing agents. We will now study effects
of heterogeneous transport parameters on these two active motion models.

\section{Superstatistics of heterogeneous populations of active particles}
\label{sec3}

The above models were established assuming a fixed diffusivity and speed for
all particles, corresponding to a homogeneous environment and population.
In contrast, many realistic systems, in which active particles are present,
are characterised by a heterogeneous environment, and populations of active
particles such as bacteria or amoeba will always involve heterogeneities in
the population, here reflected in variation of the speed $v_0$ from one
particle to the next. Assuming a distribution of diffusivities or speeds
across our population of active agents is a direct way to account for these
heterogeneities, in the spirit of the superstatistical approach \cite{beck}.
The two cases of random diffusivities and random speeds will be treated
separately for the APP and ABP models, respectively.

\subsection{Distribution of diffusivities}

The superstatistical approach with distributed diffusivity
was previously used to explain non-Gaussianity
of, inter alia, passive colloidal particles \cite{granick} as well as nematodes
\cite{hapca}. It consists in obtaining the PDF $P(\mathbf{r},t)$ of a particle
ensemble from the conditional Gaussian PDF $G(\mathbf{r},t|D)$ with a fixed
diffusivity $D$ of a single particle, through the ("superstatistical") averaging
\cite{beck}
\begin{eqnarray}
P(\mathbf{r},t)=\int_0^{\infty}p(D)G(\mathbf{r},t|D)dD, 
\label{superstat}
\end{eqnarray}
where $p(D)$ is the PDF of diffusivities characterising the system. For instance,
it can be shown that the one-dimensional exponential displacement PDF
\begin{equation}
P(x,t)=\frac{1}{\sqrt{4D_{\star}t}}\exp\left(-\frac{|x|}{\sqrt{D_{\star}
t}}\right),
\label{1d_laplace}
\end{equation}
exactly follows from an exponential form \cite{diffdiff}
\begin{equation}
\label{expd}
p(D)=\frac{1}{D_\star}\exp\left(-\frac{D}{D_\star}\right).
\end{equation}
As mentioned, such
exponential distributions for $P(x,t)$ and $p(D)$ can be found in a wide range
of systems, e.g., \cite{granick,hapca}. A possible reasoning for the occurrence
of such an exponential PDF was explored in \cite{petrovski}. Namely, writing
$D=v^2\tau/2$, where $v$ is the speed of a particle and $\tau$ the time of a
single step, one can show that when $v$ follows a Rayleigh distribution, the
PDF of $D$ is indeed exponential.

The two-dimensional PDF obtained from the diffusivity-PDF (\ref{expd})
reads \cite{diffdiff}
\begin{equation}
P(\mathbf{r},t)=\frac{1}{2\pi D_{\star}t}K_0\left(\frac{r}{\sqrt{D_\star t}}
\right),
\label{laplace_2d}
\end{equation}
where $K_0$ is the modified Bessel function of the second kind. $P(\mathbf{r}
,t)$ in \eqref{laplace_2d} is a symmetrical, bivariate Laplace distribution,
which has the general formula
\begin{equation}
\fl g(x,y,\sigma_1,\sigma_2,\rho)=\frac{1}{\pi\sigma_1\sigma_2\sqrt{1-\rho^2}}
K_0\left(\sqrt{\frac{2(x^2/\sigma_1^2-2\rho xy\sigma_1\sigma_2+y^2/\sigma_
2^2)}{1-\rho^2}}\right),
\end{equation}
where here the correlation between the two variables vanishes, $\rho=0$, and
$\sigma_1^2=\sigma_2^2=2D_{\star}t$ are the respective variances of the two random
variables $x$ and $y$. The marginal of such a bivariate symmetric Laplace
distribution is itself a Laplace distribution \cite{laplace} that can be
expressed as\footnote{The factor $\sqrt{2}$ allows us to express the
univariate Laplace distribution as a function of $\sigma_1$, the square root
of its variance.} 
\begin{equation}
g(x)=\frac{1}{\sqrt{2}\sigma_1}\exp{\left(-\sqrt{2}\frac{|x|}{\sigma_1}
\right)},
\end{equation}
which allows us to retrieve exactly the one-dimensional form \eqref{1d_laplace}
starting from \eqref{laplace_2d}. This shows that the marginal of the bivariate
symmetric Laplace distribution obtained from superstatistics with an exponential
distribution of diffusivities is the same as the PDF obtained in the
one-dimensional case from the same diffusivity-PDF $p_D(D)$.

Based on these results for the passive-diffusive case we now study what type
of displacement PDF arises from different diffusivity distributions in the
case of active particles. The superstatistical approach can again be applied
to the long time diffusive limit for active particles, knowing that
\begin{equation}
G(\mathbf{r},t|D_{\mathrm{eff}})=\frac{1}{4\pi D_{\mathrm{eff}}t}\exp{\left(
-\frac{\mathbf{r}^2}{4D_{\mathrm{eff}}t}\right)}.   
\label{active_Gauss}
\end{equation}
The dependency of the effective diffusivity on the initial diffusion
coefficients varies with the choice of the model, $D_{\mathrm{eff}}=v_0^4/
2D$ for the APP case and $D'_{\mathrm{eff}}=D_T+v_0^2/2D_R$ for the APP
model. Both cases will be studied in the following subsections.

\subsubsection{APP case.}

We study the case of an exponentially distributed diffusivity, see equation
\eqref{expd}. Combining the expression (\ref{deff}) of $D_{\mathrm{eff}}$
with equation (\ref{active_Gauss}) in (\ref{superstat}) yields
\begin{eqnarray}
\nonumber
P(\mathbf{r},t)&=&\frac{1}{D_\star}\frac{1}{2\pi v_0^4t}\int_0^{\infty}D
\exp\left(-\frac{\mathbf{r}^2D}{2v_0^4t}\right)\exp\left(-\frac{D}{D_{\star}}
\right)dD,\\
&=&\frac{1}{D_\star}\frac{1}{2v\pi_0^4t}\int_0^{\infty}D\exp{(-aD)}dD,
\label{int2_cond2}
\end{eqnarray}
with $a=\mathbf{r}^2/2v_0^4t+1/D_\star$ and leads to the Cauchy-type
PDF
\begin{equation}
P(\mathbf{r},t)=\frac{2v_0^4t}{\pi D_\star\left(\mathbf{r}^2+2v_0^4t/D_
\star\right)^2}.
\label{rad_vect}
\end{equation}
The radial PDF can readily be derived as the average $P(r,t)=\int_0^{2\pi}
P(\mathbf{r},t)rd\theta$ over the polar angle $\theta$ ($\mathbf{r}=r(\cos
\theta,\sin\theta))$, producing
\begin{equation}
P(r,t)=\frac{4rv_0^4t}{D_\star\left(r^2+2v_0^4t/D_\star\right)^2}.
\label{P_r_t}
\end{equation}
The marginal PDF $P_X(x,t)$ can be derived from (\ref{rad_vect}) using
$r^2=x^2+y^2$ and integrating out $y$,
\begin{equation}
P_X(x,t)=\frac{v_0^4t}{D_\star\left(x^2+2v_0^4t/D_\star\right)^{3/2}}.
\label{P_x_t}
\end{equation}

\begin{figure}
\centering
\includegraphics[height=6.4cm]{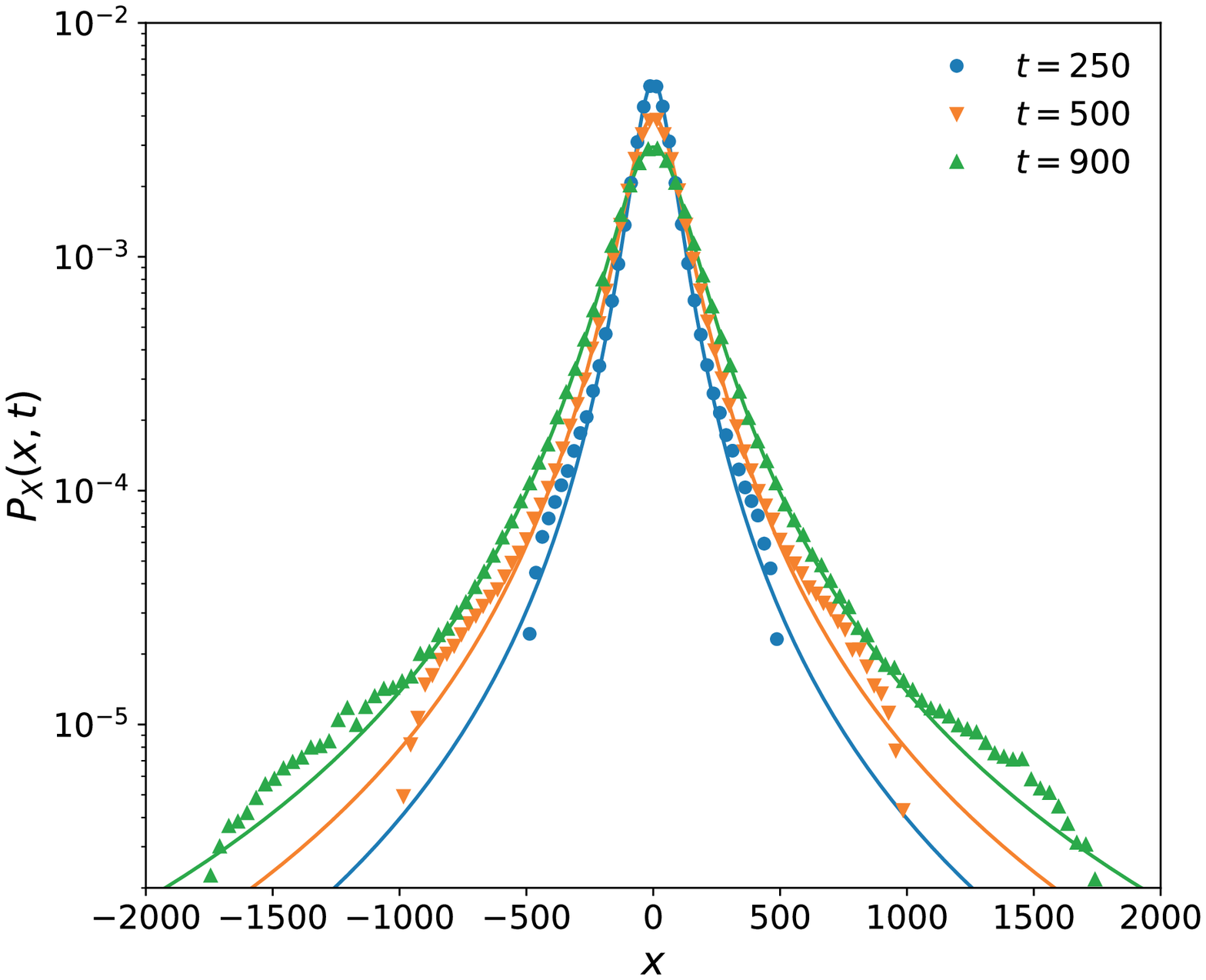}
\label{subfig:x}
\includegraphics[height=6.4cm]{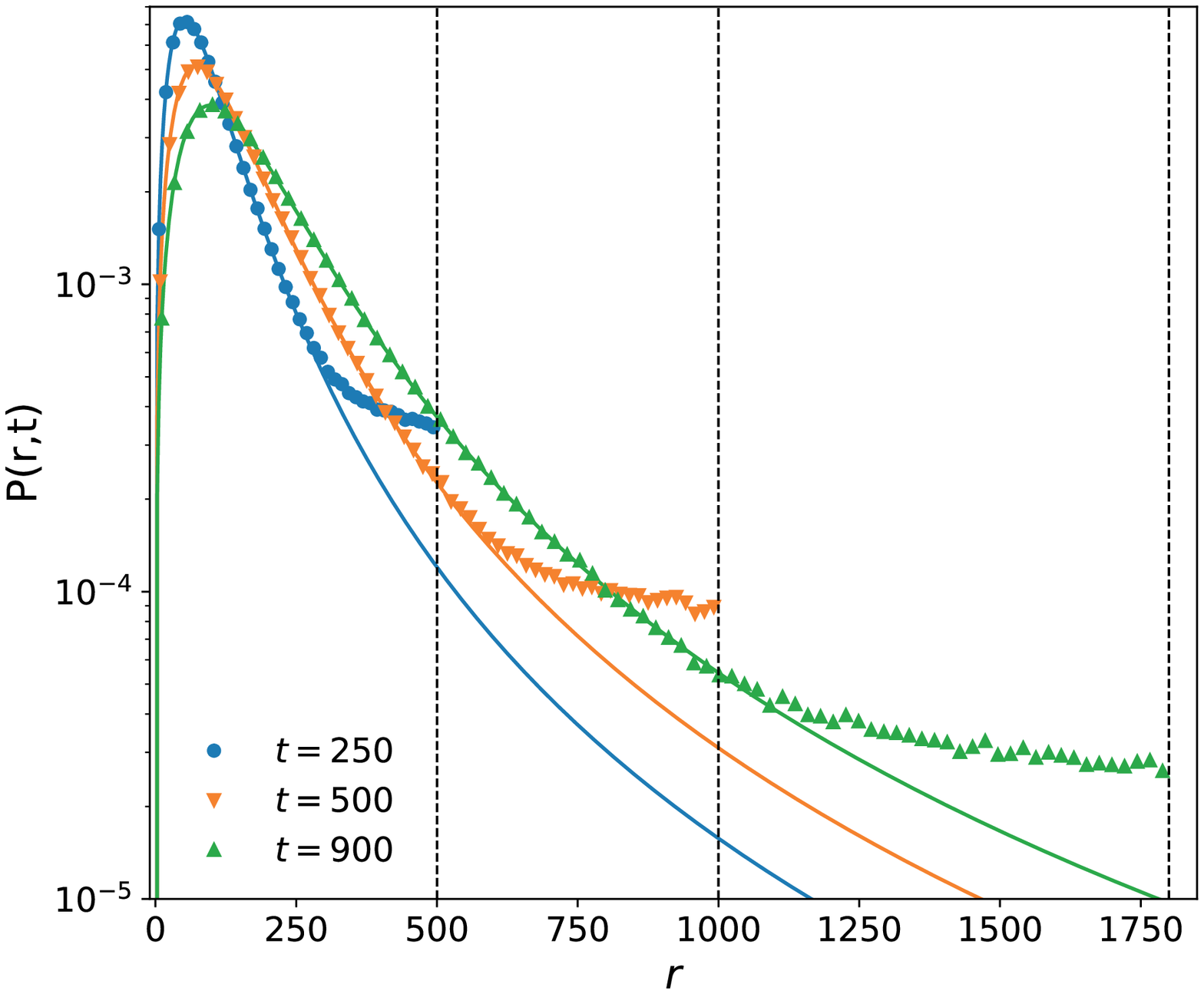}
\label{subfig:MSD}
\includegraphics[height=6cm]{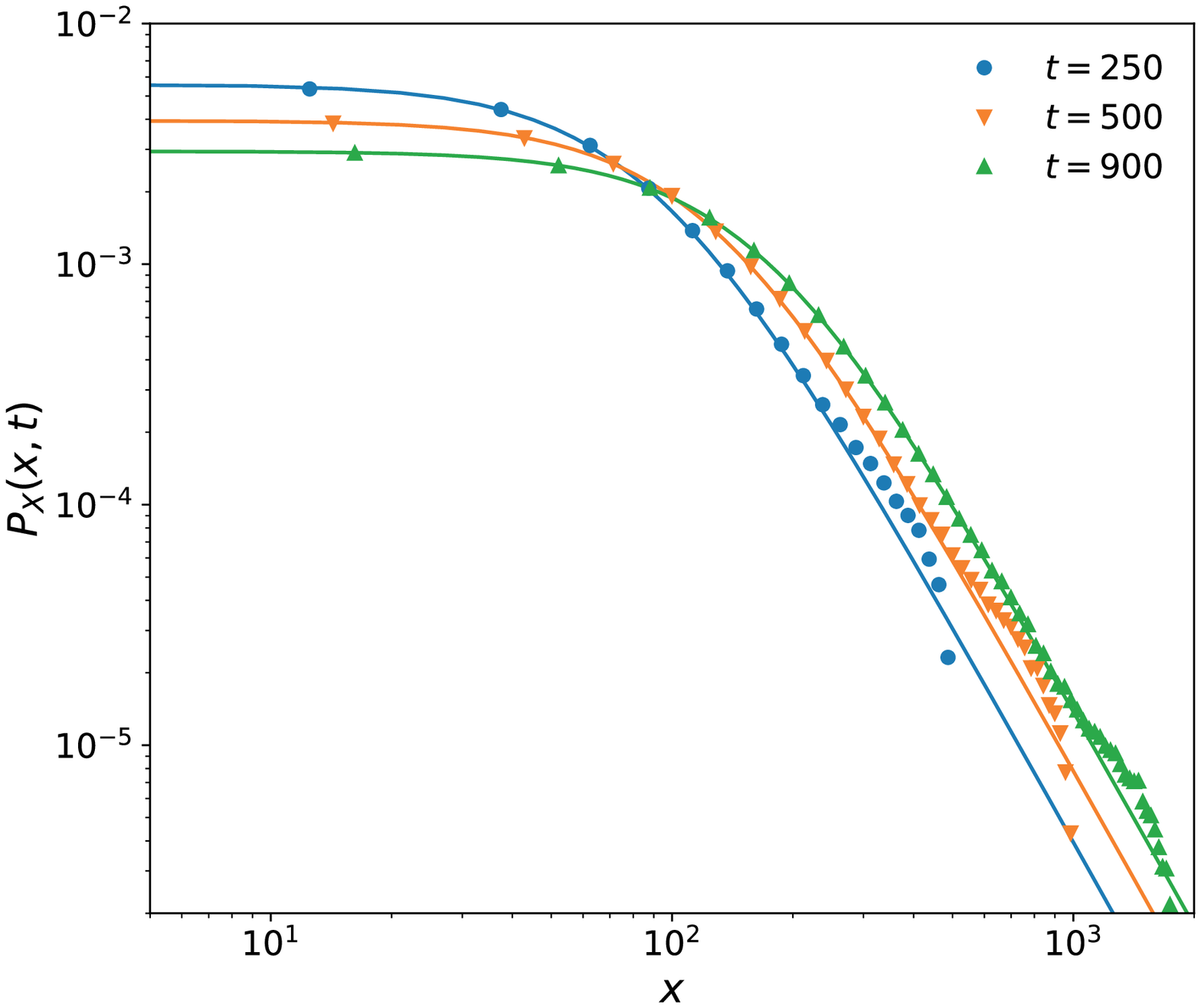}
\label{subfig:x1}
\includegraphics[height=6cm]{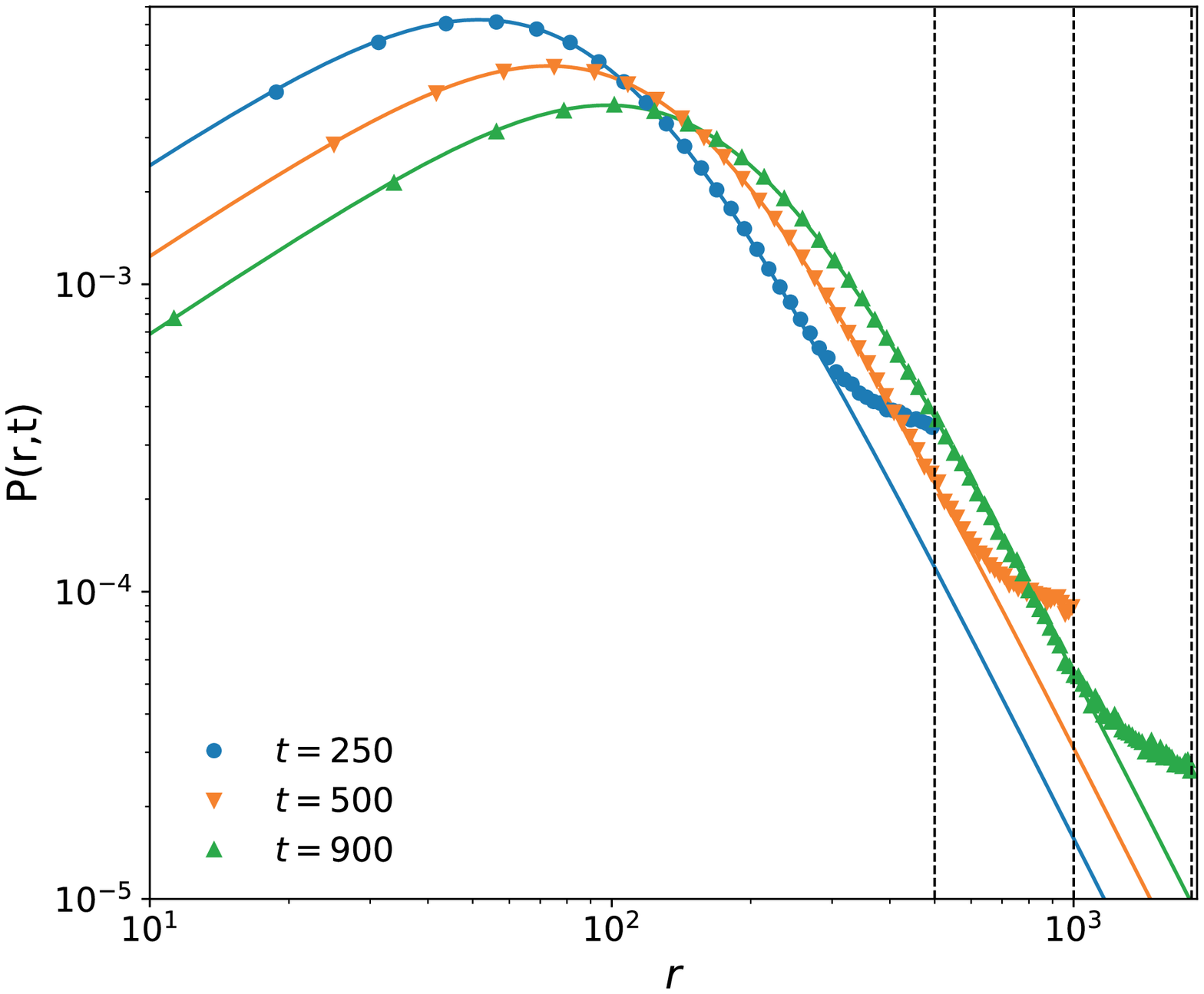}
\label{subfig:MSD1}
\caption{Superstatistical APP model. Top row: For different times we
compare the analytic solution (solid lines) with simulations results
(dots) for the marginal PDF \eqref{P_x_t} (left) and the radial PDF
\eqref{P_r_t} (right). Parameters: $v_0=2$, $D_\star=1$, with time step
$dt=10^{-2}$ and number of simulated trajectories $M=10^5$.
Bottom row: log-log plots of the same cases. The dashed lines in the
right panels indicate the maximal distance a particle with speed $v_0$
can travel without any directional changes.
\label{fig:super}}
\end{figure}

The obtained PDF $P(\mathbf{r},t)$ along with the marginal and radial PDFs
$P_X(x,t)$ and $P(r,t)$ are power-laws of finite mean and infinite higher
order moments. Stochastic simulations of the Langevin equation (\ref{vdyn})
for $M=10^5$ noise realisations were
run using an Euler-Maruyama integration scheme with a time step $\delta t=
10^{-2}$ and compared to analytics, with parameters $D_\star=1$ and $v_0=2$.
Figure \ref{fig:super} compares the marginal PDF in the $x$-direction and
the radial PDF for different times, showing very good agreement for small
to medium displacements.

Let us add a remark on the discrepancy seen for larger displacements in figure
\ref{fig:super} which is connected to the fact that the particles can travel
a maximum distance with a given speed. Indeed, if we assume a perfectly
directed motion (i.e. $d \mathbf{r}/dt=v_0\mathbf{e_{v}}$ with fixed $\phi$)
the maximum radius that can be reached at a given time is $r_{\mathrm{max}}(t)=v_0t$. This
value is shown by the black dashed lines in figure \ref{fig:super} for
the radial PDF. We see how the simulated points are limited by the front
$v_0t$, while the analytic PDF stretches much further. This discrepancy
is already present for the original APP case with fixed $D=D_\star$,
for which the long-time PDF \eqref{active_Gauss} gives rise to the radial
PDF $G(r,t)=(rD_{\star}/v_0^4t)\exp{(-D_{\star}r^2/2v_0^4t)}$, and thus
\begin{equation}
G(v_0t,t)=\frac{D_{\star}}{v_0^3}\exp{\left(-\frac{D_{\star}t}{2v_0^2}\right)}.
\end{equation}
In the present case of an an exponentially distributed diffusivity, we have
\begin{equation}
\label{pdff}
P(v_0t,t)=\frac{4v_0^5t^2}{D_\star}\frac{1}{\left(v_0^2t^2+2v_0^4t/D_\star
\right)^2}.
\end{equation}
As we are only considering $t\gg1$ we see that $P(v_0t,t)\propto1/t^2\gg G
(v_0t,t)\propto\exp{(-at)}$. Due to the much faster exponential decay in
the original APP model
the amplitude of the PDF is effectively suppressed, such that no significant
discrepancy is visible in this case. This is in contrast to the much slower
decay of the Cauchy-type PDF (\ref{pdff}) for the case of a distributed
diffusivity.\footnote{In fact, for some cases such as heat conduction, the
finite propagation speed needs to be included, as well, see
the discussion in \cite{jou}.} For short and intermediate distances, we
stress that our analytical description works very well. Note that for the
marginal PDF in the left panels of figure \ref{fig:super} the effect is
less distinct.

The first radial moment of the PDF \eqref{P_r_t} is defined as $\langle
r(t)\rangle=\int_0^{\infty}rP(r,t)dr$, and we obtain
\begin{equation}
\langle r(t)\rangle=\frac{\pi}{2}\sqrt{\frac{2v_0^4}{D_\star}}t^{1/2}.
\end{equation}
The second moment $\langle r^2(t)\rangle$ diverges, as $\lim_{r\rightarrow
\infty}r^2P(r,t)=1/r$, whose indefinite integral is the natural logarithm
and is therefore not defined. This is rather different to the analogous
superstatistical approach for passive particles, where the MSD is the same
as in the Brownian case, only the single-particle diffusivity being replaced
by the effective diffusivity of the ensemble \cite{diffdiff}, see equation
(\ref{1d_laplace}).

\begin{figure}
\centering
\includegraphics[height=6cm]{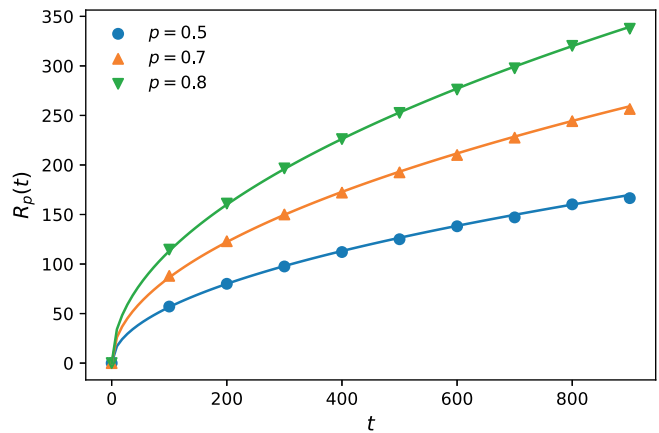}
\caption{$R_p(t)$ from numerical simulations (symbols) compared to the
analytic expression \eqref{R_p_fin} for several values of the probability
$p$. Parameters: $v_0=2$, $D_\star=1$, $dt=10^{-2}$, and number of simulated
trajectories $M=10^5$.}
\label{fig:Rp}
\end{figure}

A common way to evaluate the displacement in the case of distributions with
non-definite second moments is to introduce the quantile $p$ and the respective
horizon $R_p(t)$ such that 
\begin{equation}
\int_0^{R_p(t)}P(r,t)dr=p<1,
\label{R_p}
\end{equation}
which quantifies the advancement of the diffusion front without the need
to compute the MSD. To compute $R_p(t)$ we rewrite \eqref{R_p} such that
\begin{equation}
\int_0^{R_p(t)}\frac{D_\star}{v_0^4t\left(\frac{D_\star}{2v_0^4}\frac{r^2
}{t}+1\right)^2}rdr=\int_0^{D_\star R_p^2(t)/2v_0^4t}\frac{1}{(1+y^2)}dy=p,
\label{R_p2}
\end{equation}
where $y=D_\star r^2/(2v_0^4t)$. The result is
\begin{equation}
R_p(t)=t^{1/2}\sqrt{\left(\frac{p}{1-p}\right)\frac{2v_0^4}{D_\star}},
\label{R_p_fin}
\end{equation}
exhibiting a square root scaling of the horizon, similar to
normal diffusion and despite the infinite MSD. In figure \ref{fig:Rp} we see
excellent agreement of the analytic form \eqref{R_p_fin} for $R_p(t)$ with the
simulations.

The square-root scaling for $R_p(t)$ encountered here reminds of the results
obtained from superstatistics as applied to passive particles, where the
MSD is shown to retain a linear scaling in time \cite{hapca}, which matches
various experimental results.

\subsubsection{ABP case.}

In the ABP model both translational and rotational diffusivity coefficients
influence the particle motion. Considering purely thermal fluctuations
and spherical particles, following the Stokes-Einstein relation into
the expression (\ref{deffprime}) for $D'_{\mathrm{eff}}$ produces
$D'_{\mathrm{ eff}}=D_T+a/D_T$, with $a=2R^2v_0^2$, where $R$ is the
particle radius \cite{complex_crowded, lowen}. This case was analysed
in \cite{khadem}. Coupling the two degrees of freedom by enforcing the
Stokes-Einstein relation may unduly restrict the scope of our study
to spherical active agents exclusively undergoing passive, thermal
fluctuations. As we are dealing with a far-from-equilibrium system, often
also with non-spherical particles, we will consider the case when $D_R$
and $D_T$ are unrelated. In particular, we keep $D_T$ fixed, while $D_R$
is varied exponentially, $p(D_R)=(1/D_{R\star})\exp(-D_R/ D_{R\star})$,
across the ensemble.

The superstatistical displacement PDF for an exponential rotational
diffusivity-PDF is then generally given by the integral
\begin{eqnarray}
\nonumber
P(\mathbf{r},t)&=&\frac{1}{4\pi D_{R\star}t}\int_0^{\infty}\frac{1}{D_T
+v_0^2/2D_R}\exp\left(-\frac{\mathbf{r}^2}{4(D_T+v_0^2/2D_R)t}\right)\\
&&\times\exp\left(-\frac{D_R}{D_{R\star}}\right)dD_R.
\label{od_superstat}
\end{eqnarray}
When we assume that the translational diffusivity is much smaller than the
active propulsion term, $D_T\ll v_0^2/2D_{R\star}$, which appears a reasonable
assumption for propulsion-dominated active transport, the above integral
has the same form as \eqref{int2_cond2}, the only difference being the
specific $v_0$ dependency ($v_0^2$ instead of $v_0^4$ in \eqref{int2_cond2}).
We therefore find
\begin{equation}
P(\mathbf{r},t)=\frac{1}{D_{R\star}}\frac{1}{2\pi v_0^2t}\int_0^{\infty}
D_R\exp{(-aD_R)}dD_R,
\end{equation}
where $a=r^2/2v_0^2t + 1/D_{R\star}$. Following the same procedure as above
we then get the radial and marginal PDFs,
\begin{eqnarray}
P(r,t)&=&\frac{4rv_0^2t}{D_{R\star}\left(r^2+2v_0^2t/D_{R\star}\right)^2},
\label{P_r_t,o}\\
P_{X}(x,t)&=&\frac{v_0^2t}{D_{R\star}\left(x^2+2v_0^2t/D_{R\star}\right)
^{3/2}}.
\label{P_x_t,o}
\end{eqnarray}
The propagation of the diffusion front for $D_T=0$ is obtained analogously,
producing
\begin{equation}
R_{p}(t)=t^{1/2}\sqrt{\left(\frac{p}{1-p}\right)\frac{2v_0^2}{D_{R\star}}}.
\end{equation}

\begin{figure}
\centering
\includegraphics[height=6cm]{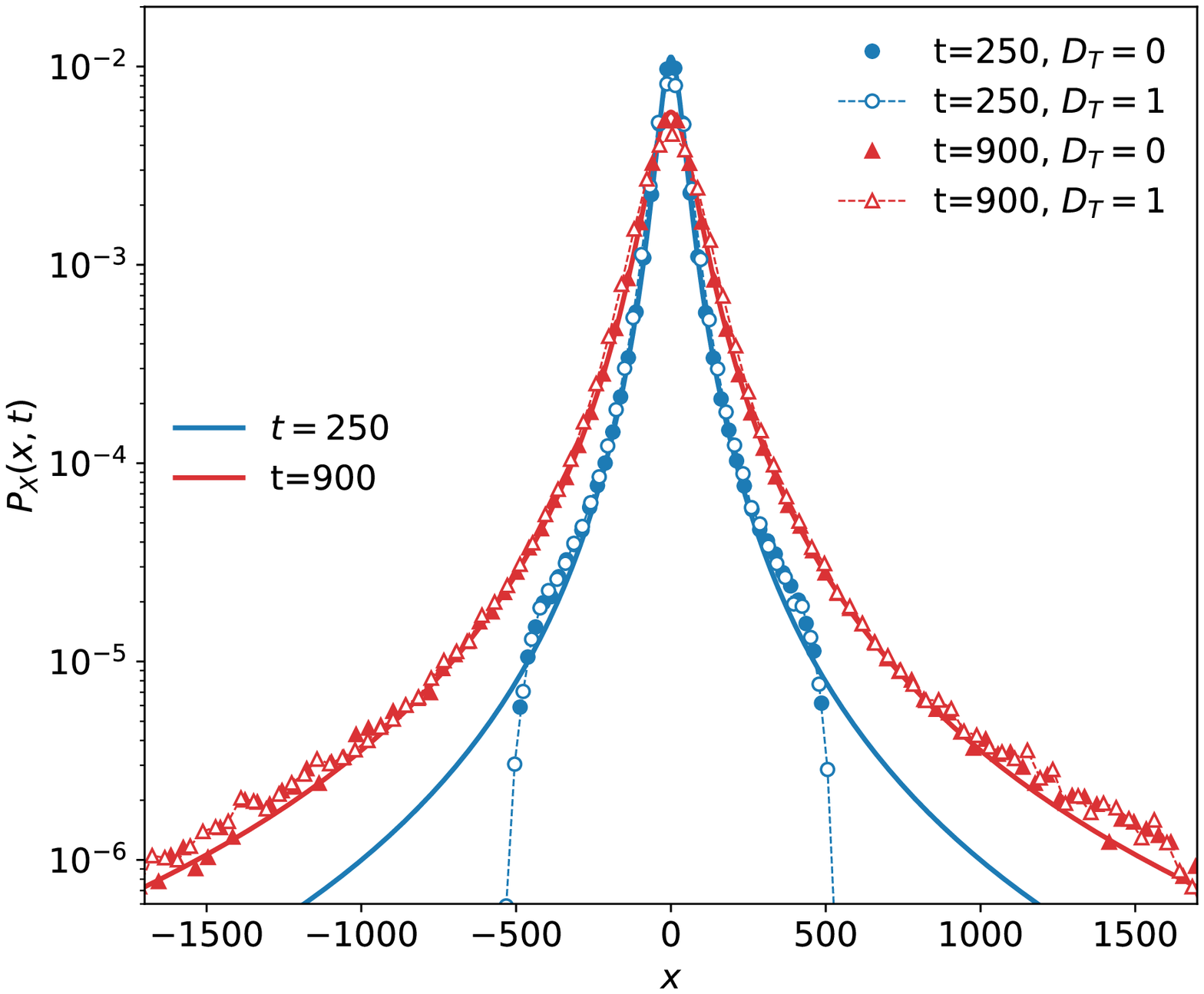}
\label{subfig:x2}
\includegraphics[height=6cm]{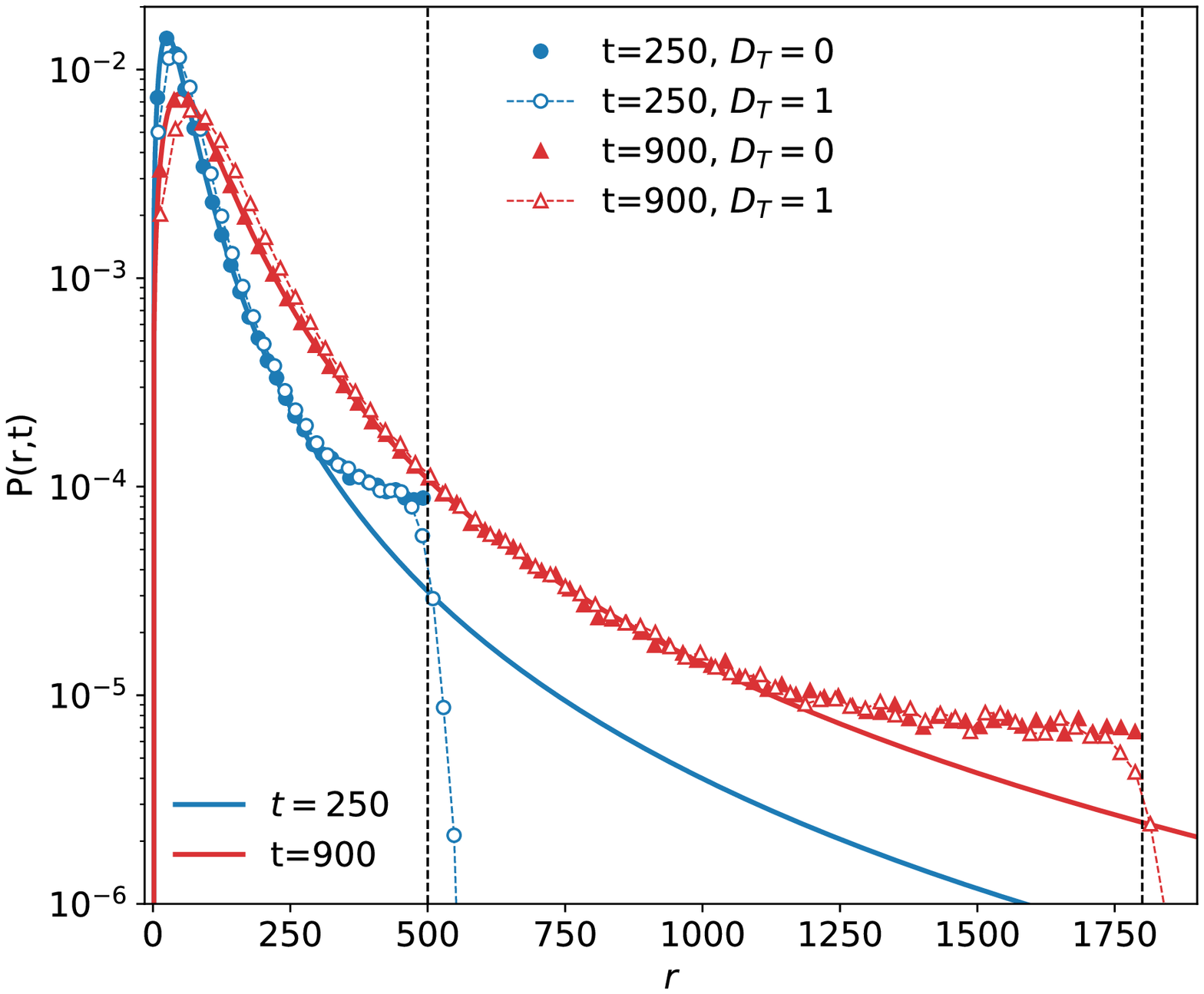}
\label{subfig:MSD2}
\includegraphics[height=6cm]{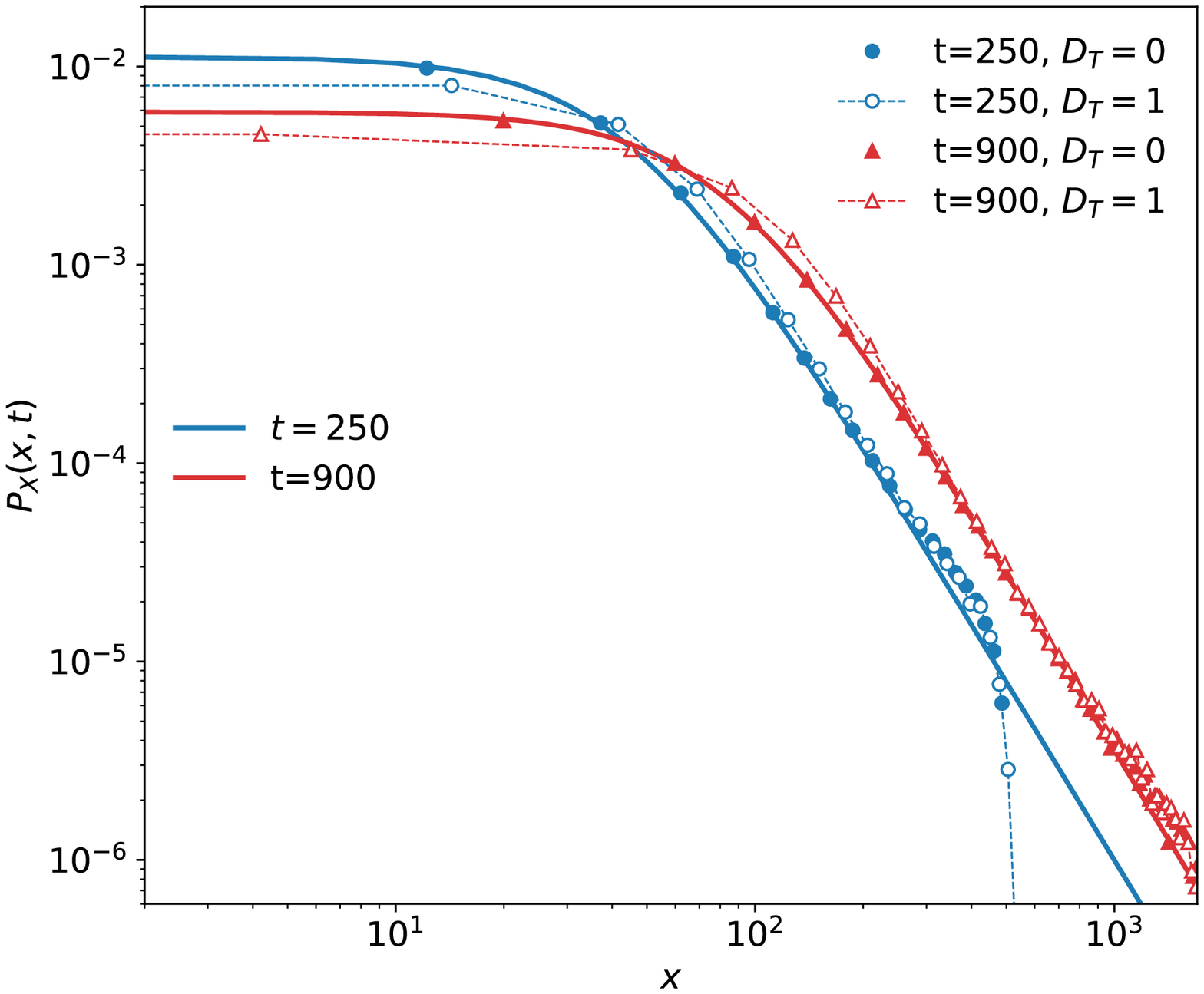}
\label{subfig:x3}
\includegraphics[height=6cm]{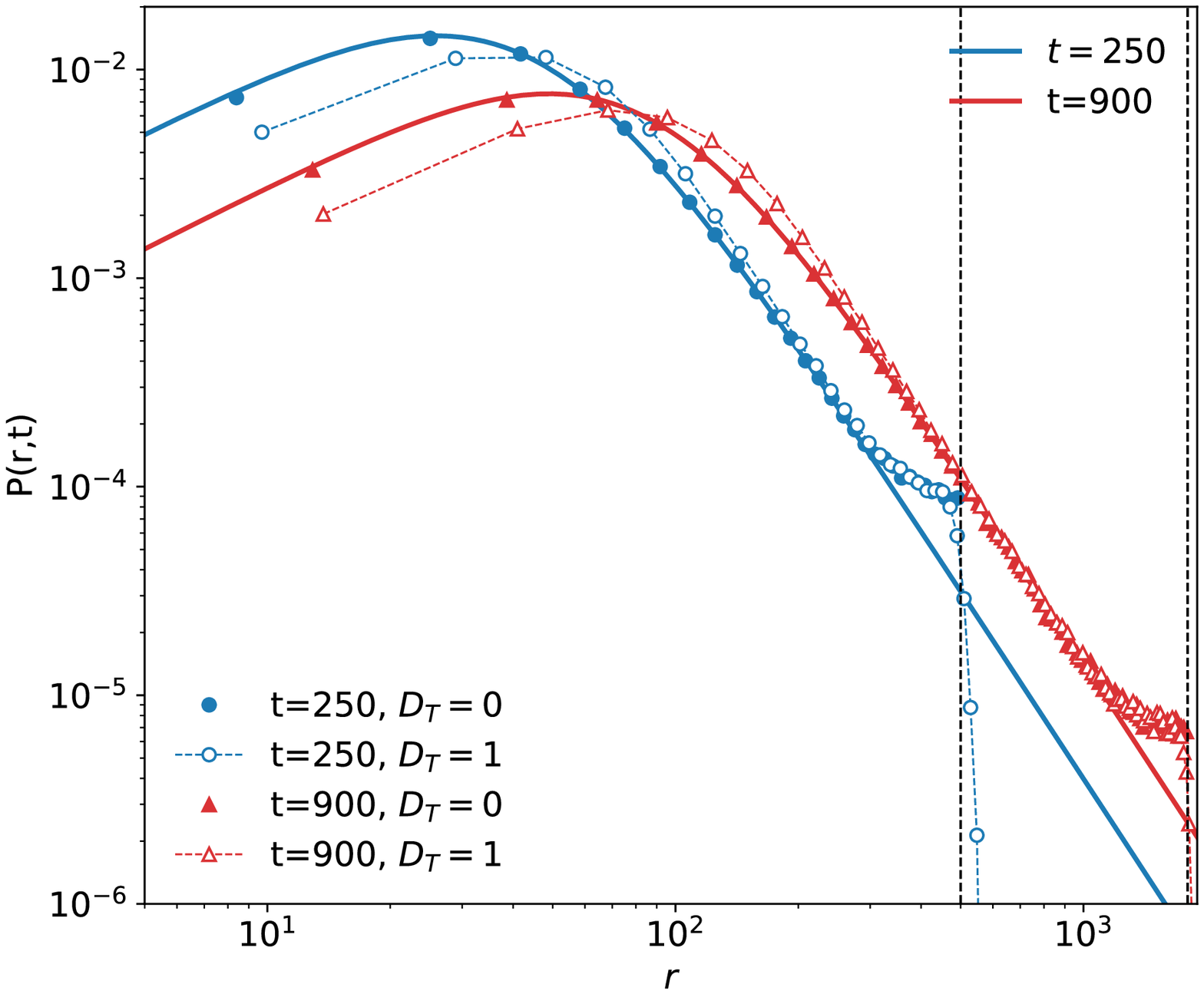}
\caption{Radial and marginal displacement PDFs for an exponential superstatistical
diffusivity-PDF in the minimal ABP model at different times. Analytical
results (solid lines) and simulations (symbols) are shown for the $x$-marginal
displacement PDF \eqref{P_x_t} (left) and for the radial PDF \eqref{P_r_t}
(right). Parameters: $v_0=2$, $D_{R\star}=1$, $\delta t=10^{-2}$, and number of
simulated trajectories $M=10^5$. The thin black line serves as a guide for
the eye. Bottom row: Same as top row but on log-log scale. Dashed vertical
lines in the right panels show the maximal distance $v_0t$ travelled
ballistically by a particle of constant speed $v_0$.}
\label{fig:super_ABP}
\end{figure}

For both marginal and radial PDFs, figure \ref{fig:super_ABP} shows the
comparison between the analytic expressions \eqref{P_r_t,o} and \eqref{P_x_t,o}
and stochastic simulations (based on equations (\ref{ActiveOverdamped1}) and
(\ref{ActiveOverdamped})) of $M=10^5$ noise realisations obtained through an
Euler-Maruyama integration scheme with a time step $\delta t=10^{-2}$. Results
from simulations including translational diffusion ($D_T=1$ in dimensionless
units) are also shown. One can observe that the shape of the PDFs remains
very similar to the Cauchy-type distributions for $D_T=0$, with a rounder
cusp for $P_{X}(x,t)$ and $P(r,t)$ peaking for slightly higher values. This
simply indicates a tendency for slightly higher displacements for higher
translational diffusivity, while the general behaviour essentially remains the
same. The cutoff at $r_{max}=v_0t$ is now less abrupt due to the presence of
finite translational diffusion.  Generally, we observe a slight shift towards
higher displacements between simulations and our approximate analytical
solution, due to the fact that we neglected the translational noise in the
analytical solutions of the ABP model. The agreement between the analytical
and numerical solutions is nevertheless good, especially for short and
intermediate $x,r$.

Cauchy-type displacement PDFs, with fat tails and infinite second moments
were therefore obtained for an exponential distribution of the rotational
diffusivities in both models of active motion discussed here. This fact
is in striking contrast to superstatistical and diffusing diffusivity
models for purely passive diffusion, in which exponential diffusivity-PDFs
effect exponential displacement PDFs (at short times for the case of
diffusing diffusivity dynamics) \cite{diffdiff}. The occurrence of
fat-tailed displacement PDFs from such a quite narrow, exponential
diffusivity-PDF is an interesting feature of the active motion models.
While in real systems such Cauchy-type shapes may be obscured by
finite size effects and other factors such as heterogeneity and/or
noise, it should be worthwhile comparing measured displacement PDFs
to these "exotic" shapes.

\subsection{Distribution of speeds}

A heterogeneous population of bacteria, amoeba, insects, etc., will always
be characterised by a variety of shapes, sizes, persistence times, and
speeds. We here consider superstatistical distributions of speeds, in an
"inverse engineering" approach. Namely, we aim at deriving a specific
speed-PDF in order to obtain an exponential displacement PDF. Specifically,
in a study of the individual motion of social amoeba of the kind
\emph{dictyostelium discoideum\/} it was shown that the displacement PDF at
longer lag times converges to an exponential shape \cite{beta}, whereas it
would be expected to converge to a Gaussian shape for the case of fixed
speeds.  In the experiment, the marginal PDFs are fitted with stretched
Gaussian functions of the shape
\begin{equation}
P(x,t)\propto\exp{(-a|x|^b)},
\label{beta1_2}
\end{equation}
where the stretching exponent $b$ is shown to decrease from around 1.2 to
a value close to unity for increasing lag times \cite{beta}. This may
indicate that the Laplace distribution represents the long-time limit of
this system and thus intrinsic properties of the \emph{dictyostelium}
population.

\subsubsection{APP case: Weibull speed-PDF.}

As a starting point we require a bivariate Laplace-shaped displacement
PDF, whose marginal in $x$-direction reads
\begin{equation}
P_X(x,t)=\frac{1}{\sqrt{4D_{\mathrm{eff}}^\star t}}\exp\left(-\frac{|x|}{
\sqrt{D_{\mathrm{eff}}^\star t}}\right),
\label{1d_laplace_Deff}
\end{equation}
which we know arises from an exponential distribution of typical value
$D_{\mathrm{eff}}^\star$ for the effective diffusivity $D_{\mathrm{eff}}$.
In the APP case we showed that $v=(2DD_{\mathrm{eff}})^{1/4}$, and
thus $p(v)=p(D_{\mathrm{eff}})|dD_{\mathrm{eff}}/dv|$ as $D_{\mathrm{eff}}$
is a monotonous function of $v>0$. Thus, requiring the exponential form
$p(D_{\mathrm{eff}})=(D_{\mathrm{eff}}^\star)^{-1}\exp(-D_{\mathrm{eff}}/
D_{\mathrm{eff}}^\star)$, we find the corresponding speed-PDF
\begin{eqnarray}
p(v)=\frac{4v^3}{2DD_{\mathrm{eff}}^\star}\exp\left(-\frac{v^4}{2DD_{
\mathrm{eff}}^\star}\right).
\label{speedpdf}
\end{eqnarray}
This form for $p(v)$ corresponds to a Weibull distribution of the general
shape\footnote{Weibull distributions
appear in extreme value statistics \cite{gumbel} and are commonly used in
a variety of fields, inter alia, distributions of wind speeds
\cite{weibull_2} or survivorship data \cite{weibull_1}.}
\begin{equation}
\label{weibull}
f_W(x;k,\lambda)=\frac{k}{\lambda}\left(\frac{x}{\lambda}\right)^{k-1}
e^{-(x/\lambda)^k},
\end{equation}
which in turn is a special case of the generalised gamma distribution
\begin{equation}
\label{gengamma}
f_G(x;k,\theta,\beta)=\frac{\beta}{\Gamma(k)\theta}\left(\frac{x}{\theta}
\right)^{k\beta-1}e^{-(x/\theta)^\beta}.
\end{equation}
Both are valid for for $x\ge0$ and vanish for $x<0$.

\begin{figure}
\centering
\includegraphics[height=5.4cm]{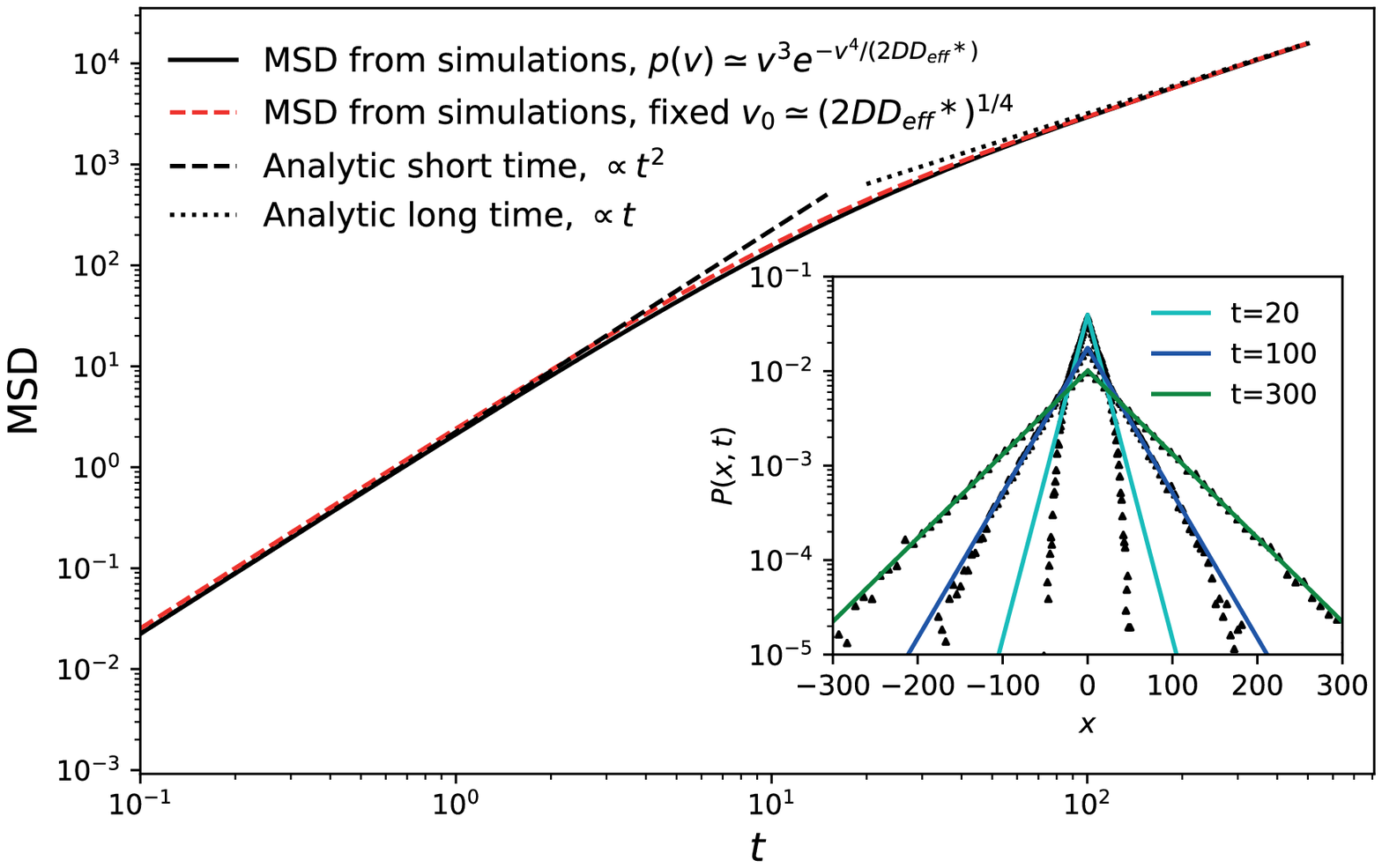}
\includegraphics[height=5.4cm]{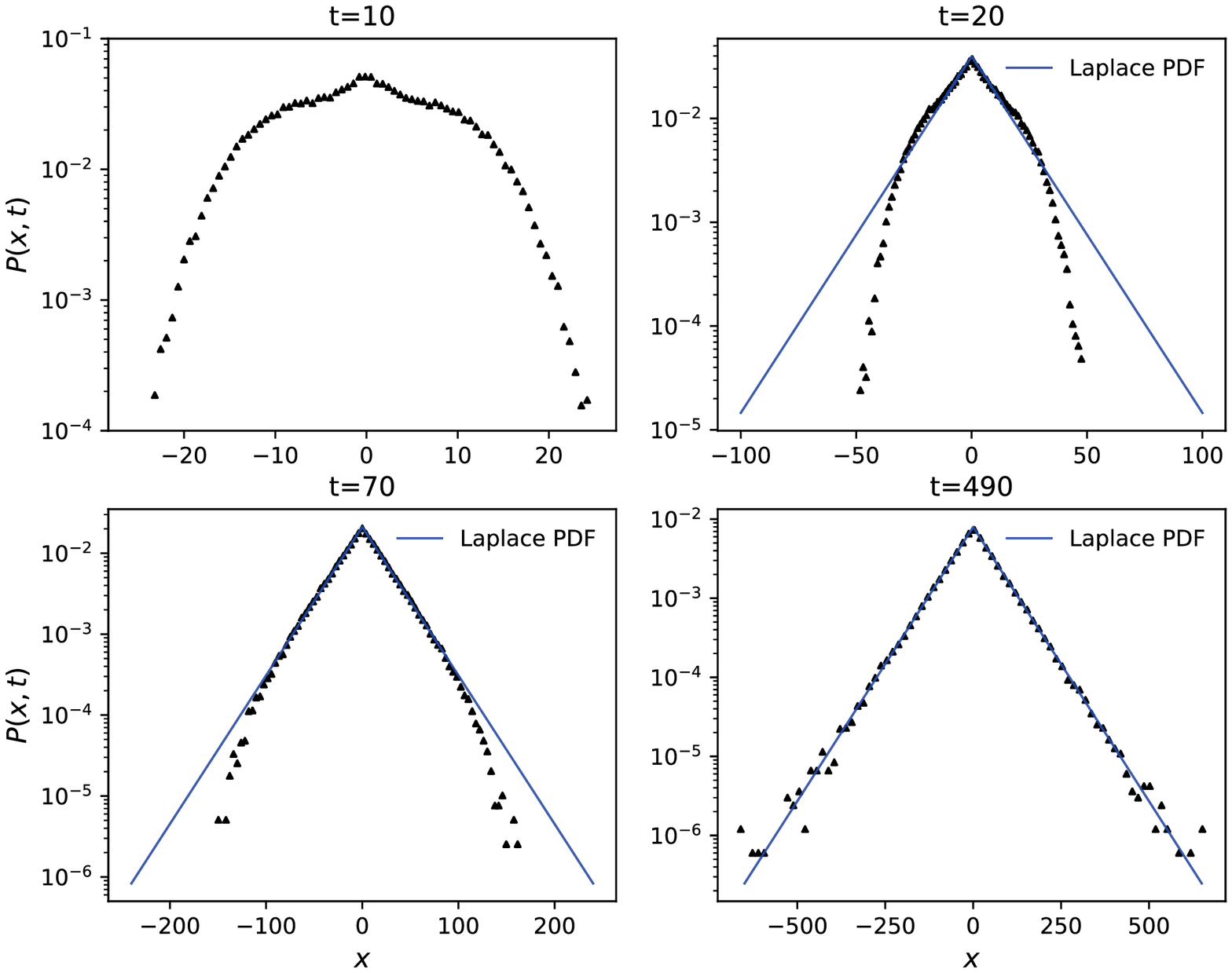}
\caption{Left: MSD and PDFs (inset) for the APP model with the Weibull-shaped
speed-PDF (\ref{speedpdf}). The MSD in the fixed-speed case
is also shown by the dashed red line. The blue, cyan, and green
solid lines (inset) indicate Laplace PDFs. Right: Marginal displacement PDF in
$x$-direction, for additional times. The solid blue lines indicate
Laplace-shaped PDFs. In both panels, we chose $D_{\mathrm{eff}}^\star=8$, $D=0.2$,
$\delta t=10^{-2}$, and the number of trajectories $M=10^5$.}
\label{fig:MSD_v0distrib_U}
\end{figure}

Stochastic simulations based on equation (\ref{vdyn}) were run with
parameters $\delta t=10^{-2}$, $D_{\mathrm{eff}}^\star=8$, and $D=0.4$, for
$M=10^5$ realisations. In figure
\ref{fig:MSD_v0distrib_U} (left) we show a comparison between the Laplace
distribution \eqref{1d_laplace_Deff} and the numerically obtained PDFs,
demonstrating very good agreement at longer time scales (we show $t=20$,
$100$, and $300$). The MSD remains exactly the same as for the fixed-speed
APP model, if we choose the propulsion speed to be $v_0=(2DD_{\mathrm{eff}}^
\star)^{1/4}$.
Thus, the MSD is ballistic at short times and linear in the long-time limit.
Additional PDFs for different times
$t$ are plotted in figure \ref{fig:MSD_v0distrib_U}
(right), where it can be seen that the PDF gradually approaches the expected
Laplace-like shape at longer times.

The generalised gamma distribution was in fact shown to describe the
speed distribution of Protozoa cells \cite{protozoa} as well as the above
mentioned \emph{dictyostelium\/} \cite{beta}. The exponents determined
from experiment are, however, quite different from those of a Weibull
distribution, with $\theta=\beta\approx1$ and $k\approx2$ resulting in
$p_{\mathrm{exp}}(v)\approx v\exp{(-v)}$. The displacement PDFs obtained
from this experimental speed distribution or from a Rayleigh distribution
(in general, from any generalised gamma distribution) can again be
derived with a similar method to that of superstatistics. Namely, using
the generalised Gamma distribution (\ref{gengamma}) for the speed and with
$v=(DD_{\mathrm{eff}})^{1/4}$, we find
\begin{equation}
\label{pexp}
p(D_{\mathrm{eff}})=\frac{\beta (2DD_{\mathrm{eff}})^{k\beta/4}}{4\theta^
{k\beta}\Gamma(k)D_{\mathrm{eff}}}\exp\left(-\left[\frac{(2DD_{\mathrm{eff}
})^{1/4}}{\theta}\right]^\beta\right).
\end{equation}
We then insert this form into the superstatistical formulation
\begin{equation}
P(\mathbf{r},t)=\frac{1}{4\pi t}\int_0^{\infty}\frac{1}{D_{\mathrm{eff}}}
p(D_{\mathrm{eff}})\exp\left({-\frac{\mathbf{r}^2}{4D_{\mathrm{eff}}t}}\right)
dD_{\mathrm{eff}}.
\label{NumerInt1}
\end{equation}
This integral can be solved using Fox $H$-functions (see \ref{appfox}),
leading to the expression
\begin{equation}
P(\mathbf{r},t)=\frac{1}{4\pi t}\frac{1}{\Gamma(k)}\frac{(2D)^{1-4/\beta}}{
\theta^{4-\beta}}H_{0,2}^{2,0}\left[\frac{2D\mathbf{r}^2}{\theta^44t}\left|
\begin{array}{l}\rule{1.2cm}{0.01cm}\\(k-4/\beta,4\beta),(0,1)\end{array}
\right.\right].
\label{before_last}
\end{equation}
For the marginal PDF in $x$, we find
\begin{equation}
P_X(x,t)=\frac{a^{l-1}}{\sqrt{4\pi t}\Gamma(k)}H_{0,2}^{2,0}\left[b\left|
\begin{array}{l}\rule{1.2cm}{0.01cm}\\(k-l+\frac{1}{2},l),(0,1)\end{array}
\right.\right]
\label{before_last_marg}
\end{equation}

\begin{figure}
\centering
\includegraphics[width=0.7\textwidth]{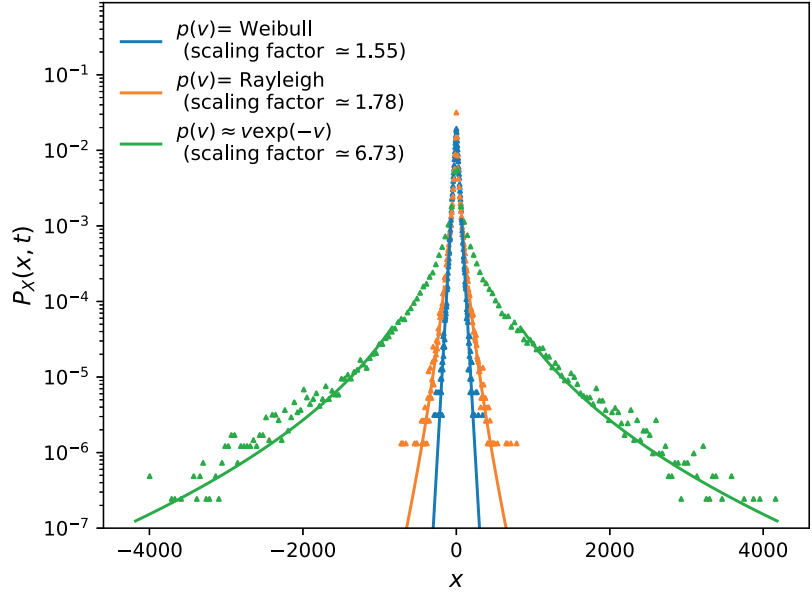}
\caption{Marginal PDF $P_X(x,t)$ for various speed PDFs in the APP model at
time $t=490$. Analytic large-displacement asymptotics as obtained from
(\ref{marg_fin}) (solid lines) and comparison with stochastic simulations
(triangles) with various distributions of the
speed $v_0$. For all distributions, $D=0.4$, the number of trajectories
is $M=10^5$, $\delta t=10^{-2}$. Based on equation (\ref{pexp}) we use
the fitted distributions from \cite{beta} with $k=2$, $\theta=\beta=1$;
a Rayleigh distribution with $k=1$, $\beta=2$, $\theta=1$; and a Weibull
distribution with $k=1$, $\beta=4$, $\theta=1$. The "scaling factor"
listed in the figure is needed to account for the constant factor, that
is unknown in the expansion (\ref{marg_fin}).}
\label{fig:APP_asympt}
\end{figure}

The large-displacement asymptote of the marginal $P_X(x,t)$ can be obtained
explicitly, yielding
\begin{equation}
\fl P_X(x,t)\simeq\frac{a^{l-1}}{\sqrt{4\pi t}\Gamma(k)}\left(\frac{a^lx^2}{4t}
\right)^{(k-l)/(1+l)}\exp\left(-\frac{1+l}{l^{l/(1+l)}}\left(\frac{a^lx^2}{
4t}\right)^{1/(1+l)}\right),
\label{marg_fin}
\end{equation}
for $(a^lx^2/4t)\to\infty$,
with $a=(2D)^{\beta/4}/\theta^{\beta}$ and $l=\beta/4$, and where
$\theta$, and $\beta$ are the parameters of the Gamma distribution.
Here the symbol $\simeq$ signifies that the expansion is known up to a
("scaling") factor of the order of unity. Additional
details of the derivation are presented in \ref{appfox}. Figure
\ref{fig:APP_asympt} compares the PDFs obtained from simulations with the
analytic asymptotic results. In particular, we see that the effective
diffusivity-PDF corresponding to the the experimentally observed speed PDF
$p(v)\approx v\exp(-v)$ in \cite{beta} (green line), produces a marginal PDF
with a sharper cusp and significantly broader tails than a Laplace PDF. We
will show that the sought-after consistency between the experimental speed-
and displacement-PDFs can be recovered in the ABP model.

\subsubsection{ABP case: Rayleigh PDF.}

\begin{figure}
\centering
\includegraphics[width=0.6\textwidth]{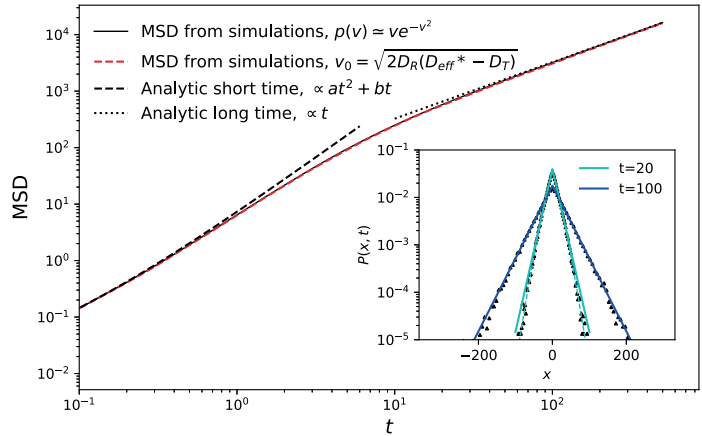}
\caption{MSD and PDFs for the ABP model in the case of a Rayleigh speed-PDF.
The MSD in the case with fixed speed $v_0=\langle v\rangle$ equalling the
mean of the speed of the Rayleigh PDF is also shown (dashed red
line). The solid lines (inset) indicate exponential PDFs, while the dashed
line is a best fit to the shape $A\exp{(|x|^{\alpha}/b)}$, $\alpha \simeq
1.21$. Here $D_{\mathrm{eff}}^\star=8$, $D_R=0.4$, $D_T=0.2$, and $\delta
t=10^{-2}$, for $M=10^5$ trajectories.}
\label{fig:MSD_v0distrib_O}
\end{figure}

\begin{figure}
\centering
\includegraphics[height=6cm]{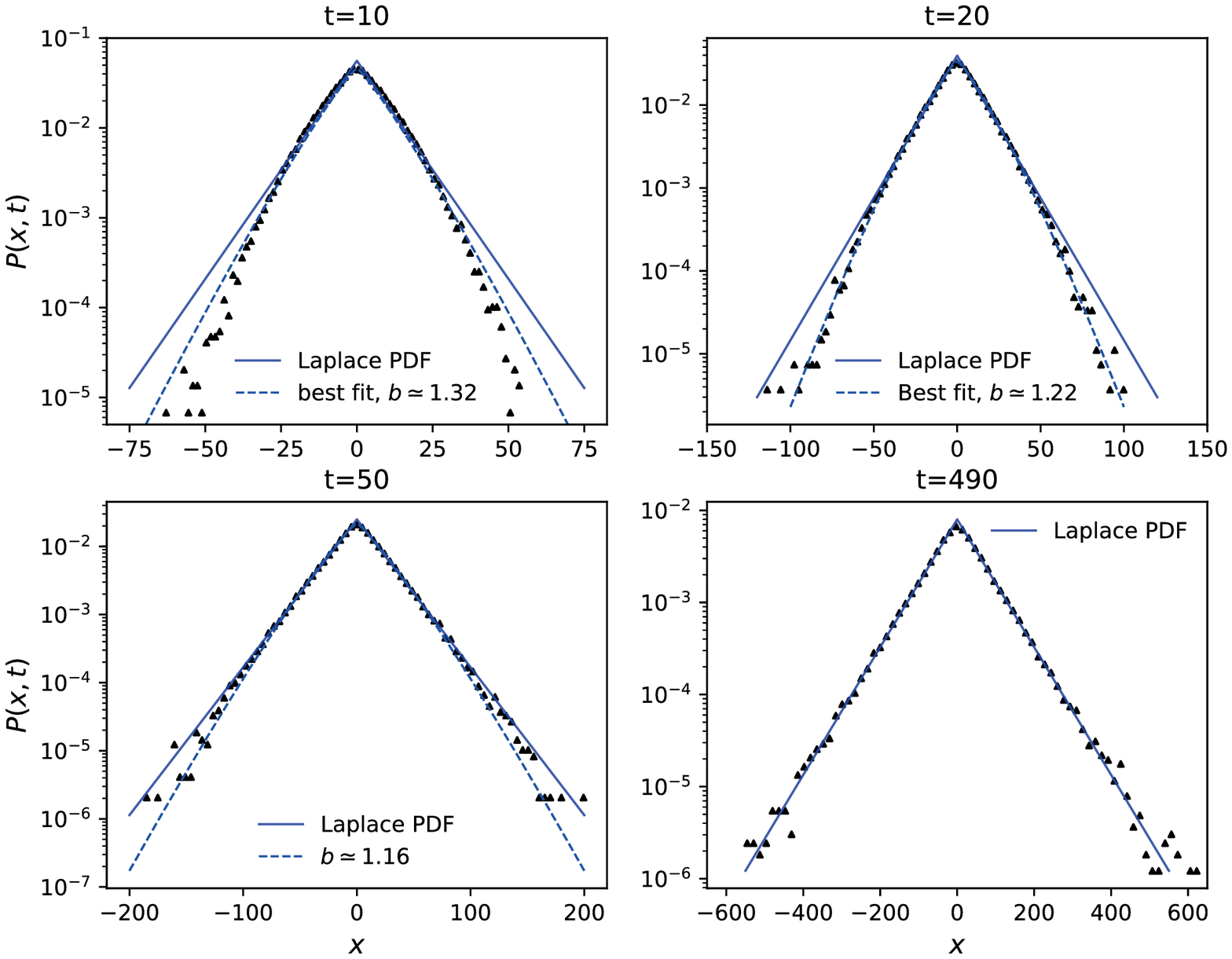}
\label{subfig:ray}
\includegraphics[height=6cm]{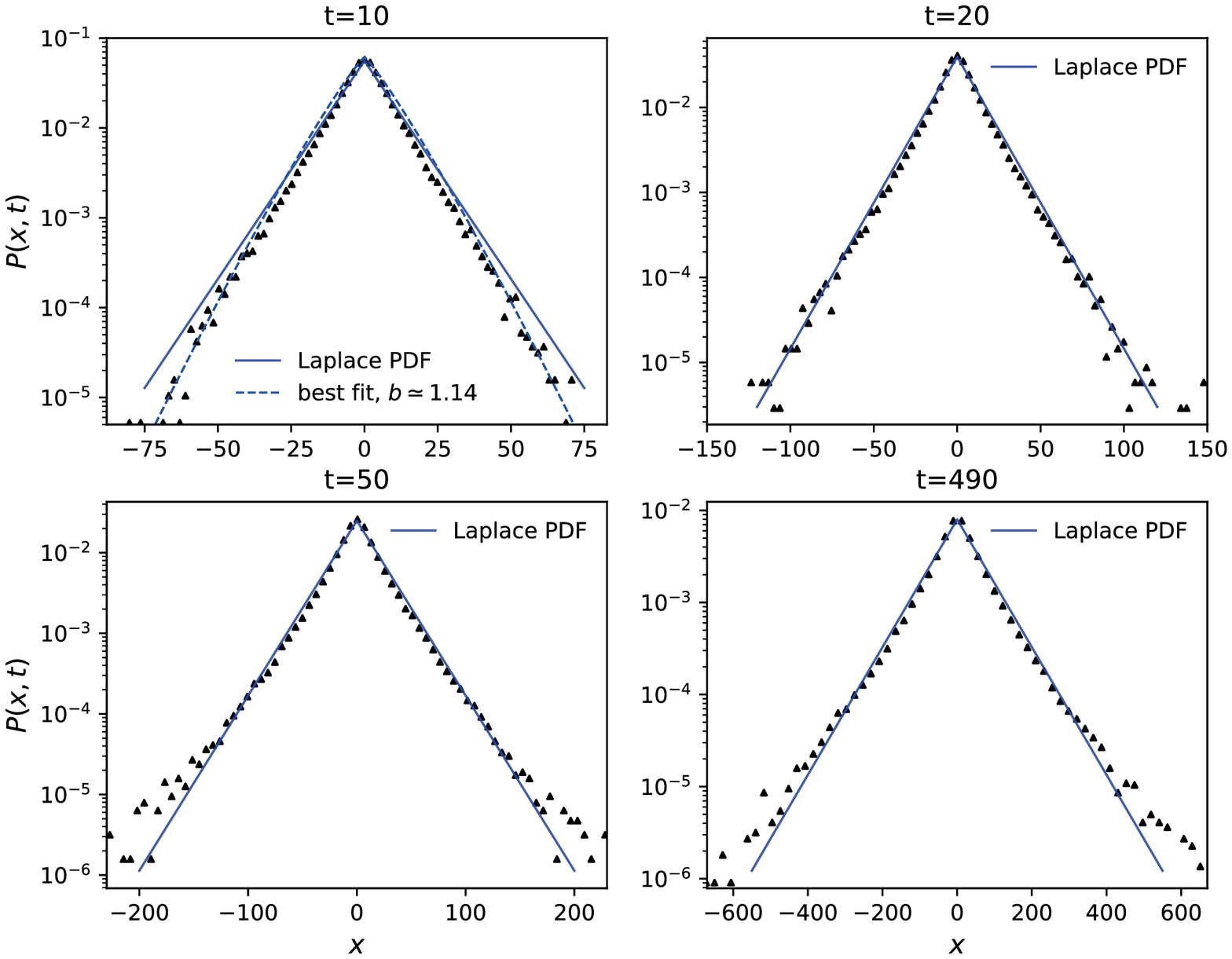}
\label{subfig:exp}
\caption{Marginal displacement PDFs in $x$-direction from simulations in
the ABP model for a Rayleigh speed-PDF (left) and the experimentally
fitted speed-PDF (right). The solid lines indicate an exponential PDF
while the dashed lines are best fits for functions of the type $A\exp(-
|x|^b/c)$. The parameters are $D_{\mathrm{eff}}^\star=8$, $D_R=0.4$,
$D_T=0.2$, $\delta t=0.01$, and the number of trajectories is $M=10^5$.}
\label{fig:x_distribs_O}
\end{figure}

We now assume that the long-time diffusivity $D'_{\mathrm{eff}}=D_T+v_0^2/2
D_R$ of the ABP model follows the exponential distribution $p(D'_{\mathrm{
eff}})\approx(1/D_{\mathrm{eff}}^\star)\exp{(-D'_{\mathrm{eff}}/D_{\mathrm{eff}}
^\star)}$, such that the speed-PDF has the Rayleigh shape\footnote{Here the
$\approx$ sign is due to a numerical factor $A=\exp{(D_T/D_{\mathrm{eff}}^
\star)}$ in the PDF of $D_{\mathrm{eff}}$, which is strictly defined to
range between $D_T$ and $+\infty$, and not in between $0$ and $+\infty$.
This numerical factor is, however, very close to unity with our usual choice
of parameters.}
\begin{equation}
p(v)\approx\frac{v}{D_RD_{\mathrm{eff}}^\star}\exp{\left(\frac{-v^2}{2D_RD_{
\mathrm{eff}}^\star}\right)}.
\label{vdistrib_o}
\end{equation}
This form also corresponds to the Maxwellian distribution of velocities, that
was found by Okubo and Chiang \cite{okubo} for the speed distribution of
flying insects. We compare this result to stochastic simulations of equations
(\ref{ActiveOverdamped1}) and (\ref{ActiveOverdamped}) with
parameters $\delta t=10^{-2}$, $D_{\mathrm{eff}}^\star=8$, $D_R=0.4$,
and $D_T=0.2$, for $M=10^5$ noise realisations in figure
\ref{fig:MSD_v0distrib_O}. The MSD obtained from simulations with this
initial distribution remains unchanged from the fixed-$v$ case. The PDFs
shown in figure \ref{fig:MSD_v0distrib_O} (additional times are shown in figure
\ref{fig:x_distribs_O}, left) evolve from having a close-to-Gaussian shape
at short lag times to a distribution that nicely matches the Laplace
distribution \eqref{1d_laplace_Deff} at long lag times. This corresponds
quite well to the experimental findings in \cite{beta}, where the fitted
stretching exponent $b$ (compare \eqref{beta1_2}) decreases from $b\approx
1.2$ to a value close to unity at longer lag times.

\setcounter{footnote}{1}

\begin{figure}
\centering
\includegraphics[width=0.7\textwidth]{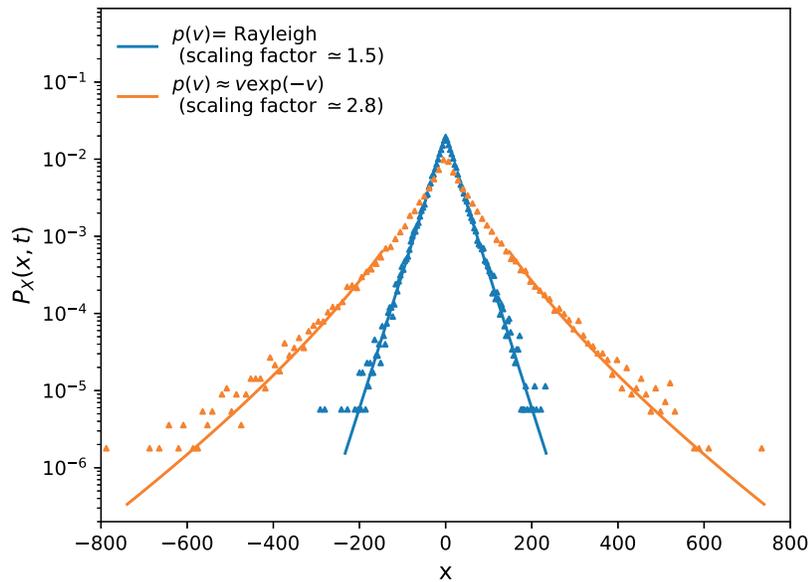}
\caption{Analytic large-displacement asymptotics for the ABP model as obtained
from (\ref{Px_ABP_paper}) (orange line) and (\ref{Px_ABP_ray}) (blue line)
and comparison with numerics (triangles),
for two PDFs of the initial speed (both with $D_R=0.4$, $D_T=0$, number of
trajectories $M=10^5$, and $\delta t=10^{-2}$). PDF with $k=2$, $\theta=
\beta=1$; Rayleigh distribution with $k=1$, $\beta=2$, $\theta=1$.
The "scaling factor"
listed in the figure is needed to account for the constant factor, that
is unknown in the expansion (\ref{marg_fin}).}
\label{fig:ABP_asymptot}
\end{figure}

In the experimental study of \emph{dictyostelium} \cite{beta} it was,
however, shown that the experimental speed-PDF grew faster at small
speeds and had more pronounced tails at high speeds. Instead of the
Rayleigh distribution the generalised two-parameter Gamma distribution
\begin{equation}
\label{pexp1}
p_{\mathrm{exp}}(v)=\frac{\gamma}{\Gamma(\varepsilon+1/\gamma)}v^{\varepsilon
\gamma}\exp\left(-v^{\gamma}\right)
\end{equation}
with $\gamma\approx1$ and $\varepsilon\approx1$, i.e., an approximately
exponential speed-PDF, was shown to provide an excellent fit. The analytic
expression of the large-displacement asymptotes for $P(\mathbf{r},t)$ and
$P_X(x,t)$ can again be obtained via Fox $H$-functions, see
\ref{appfox}\footnote{Here and in the following we set $D_T\approx0$ for
simplicity in our computations, as the effects of the translational
diffusivity are negligible for our choice of parameters, see also above.}.
In particular, the marginal PDF corresponding to the speed PDF (\ref{pexp1})
fitted to the experimental data in \cite{beta}) can, in the asymptotic limit,
be written as
\begin{equation}
P_X(x,t)\simeq\frac{1}{\sqrt{4\pi t}}\frac{1}{\Gamma(2)}\frac{\sqrt{2D_R}}{
\theta}\exp{\left(-3\times2^{-2/3}\left(\frac{\sqrt{2D_R}x^2}{4t\theta}\right)
^{1/3}\right)},
\label{Px_ABP_paper}
\end{equation}
for $D_Rx^2/(2t\theta^2)\to\infty$.
Figure \ref{fig:ABP_asymptot} shows good agreement between the simulations
and the analytic large-displacement asymptotes. We see how this speed PDF
actually leads to the asymptote $P(x,t)\propto\exp{(-ax^{2/3})}$. The
comparison between the displacement PDF obtained from simulations with this
speed PDF (\ref{pexp1}) and a Laplace PDF nonetheless shows that for intermediate
times, both distributions actually match very well.

We point out the almost perfect agreement between the marginal displacement
PDFs shown for the lag time $t=20$, at which the MSD crosses over between
the short-time ballistic and the linear long-time regime, as seen in figure
\ref{fig:MSD_v0distrib_O}.
We also note that the displacement PDFs obtained at shorter lag times
can also be fitted rather well with functions of the form \eqref{beta1_2},
with $1<b<2$, despite some difference at smaller displacements. Despite
the simple assumptions of our model in combination with the experimental
speed-PDF the agreement with the experimentally observed displacement PDF
at various times is remarkable and demonstrates that the ABP model provides
an adequate model for the \emph{dictyostelium\/} motion based on the current
data. In particular, the agreement is much better than for the APP model.

\section{Conclusion}
\label{sec4}

Common models for active stochastic motion are characterised by some
specific correlation time, beyond which the motion converges to
normal Brownian motion with a Gaussian displacement PDF and an MSD,
that is linear in time and has an effective diffusion coefficient, in
which the typical particle speed and the correlation time enter.
Following reports of non-Gaussian displacement PDFs, we here developed
superstatistical extensions of two active particle models: One is the
APP model including passive-noise. This model is mainly useful for
the description of particles with small damping coefficients, for instance,
insects or birds. The other is the ABP model with active fluctuations. For
both models we considered distributions in their diffusion coefficients
and speeds. For living creatures such variations can be thought of as
physiologically given, such that random parameters occur even in
an homogeneous environment. In this sense the phenomenon of non-Gaussian
active particles is similar to "ensembles" of non-ideal, passive tracer
particles with a significant distribution of particle sizes or shapes.

For distributed diffusivities we found that an exponential distribution
of particle diffusivities in the APP model give rise to a Cauchy-type
displacement PDF, for which we also calculated the diffusion fronts. A
similar result was obtained for the ABP case with an exponential diffusivity-PDF.
Thus, active particle models behave very differently from passive
particles, for which an exponential diffusivity PDF effects exponential
tails of the displacement PDF. For the analysis of the distribution of
speeds we required that the asymptotic displacement PDF follows a
Laplace PDF. To achieve this, in the APP model we employed a Weibull
speed-PDF. We also presented the results for a more versatile
generalised gamma distribution. For the ABP model we discussed a
Rayleigh speed-distribution and a generalised gamma distribution. It
turns out that the description in terms of the superstatistical ABP
model with random speeds selected from the experimentally measured speed-PDF
provides a very nice description of the experimentally determined marginal
displacement PDF in $x$-direction. The superstatistical ABP model driven by
active fluctuations thus provides a consistent link between the experimentally
determined displacement- and speed-PDFs.

The diffusion of active particles in evolving heterogeneous environments was
recently investigated in \cite{khadem} through the implementation of the
"diffusing diffusivity" mechanism, according to which the diffusivity of each
particle is taken as a random variable evolving through time as a stationary
stochastic process, e.g., the Ornstein-Uhlenbeck process. The obtained results
include a non-Gaussian displacement PDF
at time scales shorter than the correlation time of the auxiliary
diffusivity process. At long times, this model shows a convergence to a Gaussian
profile with an effective diffusivity. It is thus different from our
superstatistical model, in which, in an heterogeneous environment, the
parameter randomness is due to individual physiology, and the non-Gaussianity
persists in the long time limit. The latter case corresponds to the
experimental results for \emph{dictyostelium\/} \cite{beta} and
nematodes \cite{hapca}.

Active baths, i.e., an environment made up of a great number of active
particles, are of interest for both their role in concrete biological
or engineering systems and their physically remarkable behaviour due
to co-operativity effects \cite{activebath1}. Such experiments found
interesting diffusion behaviour for the passive particles, some of them
exhibiting exponential tails \cite{activebath2}. It will be interesting
to study the emerging co-operative dynamics of baths of non-identical
particles with superstatistically distributed parameters.

Following the favourable comparison of our results with \emph{dictyostelium\/}
cells analysed here, it could be interesting to see how other motile-cellular
systems behave statistically, e.g., \emph{acanthamoeba\/} cells, whose
intracellular motion in supercrowded cells was found to be strongly
superdiffusive \cite{selhuber}, and who also perform superdiffusive active
motion of the entire cell \cite{glebprx}. We could also think of applying
such modelling to larger animals, such as the motion of birds such as kites
and storks \cite{ohad}. Finally, extensions to intermittent dynamics such
as run-and-tumble models are possible \cite{thiel}.

\ack

We acknowledge support from the German Science Foundation (DFG, grant
no.~ME 1535/16-1). AC acknowledges the support of the Polish National
Agency for Academic Exchange (NAWA).

\begin{appendix}

\section{Probability density function of the director angle $\phi(t)$}
\label{letphi}

We here derive the PDF for the director angle $\phi(t)$ needed to determine
the MSD in \ref{appc}.
Let $\phi$ be a $2\pi$-periodic Wiener process that follows the Langevin equation
\begin{equation}
\frac{d\phi(t)}{dt}=\frac{1}{v_0}\sqrt{2D}\xi(t),
\label{phi_eq2}
\end{equation}
or, equivalently, the Fokker-Planck equation
\begin{equation}
\frac{\partial P(\phi,t|\phi_0,t_0)}{\partial t}=\frac{D}{v_0^2}
\frac{\partial^2P(\phi,t|\phi_0,t_0)}{\partial^2\phi}
\label{FP}. 
\end{equation}
Let us now retrieve its
PDF $P(\phi,t|\phi_0)$ for a given initial condition $\phi_0$. We use
$a=D/v_0^2$ in the following, and set $t_0=0$. One knows that $P(\phi,
t=0)=\delta(\phi-\phi_0)$ and that $0\leq\phi\leq2\pi$. We can write the
expansion 
\begin{equation}
P(\phi,t|\phi_0)=\sum_{n=-\infty}^{\infty}V_n(t)e^{in\phi}
\label{expansion} ~,
\end{equation}
and therefore know that
\begin{equation}
\delta(\phi-\phi_0)=\sum_{n=-\infty}^{\infty}V_n(0)e^{in\phi}
\label{expansion0}.
\end{equation}
Multiplying on both sides of equation \eqref{expansion0} with $\int_0^{2
\phi}d\phi e^{-im\phi}$, $m\in\mathbb{Z}$, we eventually get that
\begin{equation}
\int_0^{2\phi}\delta(\phi-\phi_0)e^{-im\phi}d\phi=\sum_{n=-\infty}^{
\infty}V_n(0)\int_0^{2\phi}e^{i(n-m)\phi}d\phi,
\label{bla}
\end{equation}
where the right-hand term is non-zero only for $n=m$. Thus,
\begin{equation}
V_n(0)=\frac{e^{-in\phi_0}}{2\pi}.
\label{v0}
\end{equation}

Now we can plug expression \eqref{expansion} into \eqref{FP} and get a
differential equations for the coefficients $V_n(t)$, such that 
\begin{equation}
\sum_{n=-\infty}^{\infty}\frac{dV_n}{dt}e^{in\phi}=- a\sum_{n=-\infty}
^{\infty}V_n(t)e^{in\phi},
\end{equation}
which yields, after repeating the procedure leading to equation \eqref{bla},
\begin{equation}
2\pi\frac{dV_m}{dt}=-2\pi aV_m(t)m^2.
\label{diff_vn}
\end{equation}
The differential equation \eqref{diff_vn} is easily solved, and we get
\begin{equation}
V_n(t)=V_n(0)e^{-n^2at}.
\label{vn} 
\end{equation}
Knowing $V_n(0)$, we can thus retrieve the expression $P(\phi,t|\phi_0)$,
\begin{numparts}
\begin{eqnarray}
\fl P(\phi,t|\phi_0)&=&\frac{1}{2\pi}\sum_{n=-\infty}^{\infty}e^{(in(\phi-
\phi_0))}e^{-n^2at},\\
\fl&=&\frac{1}{2\pi}+\frac{1}{2\pi}\sum_{n=-\infty}^{-1}e^{in(\phi-\phi_0)}
e^{-n^2at}+\frac{1}{2\pi}\sum_{n=1}^{\infty}e^{(in(\phi-\phi_0))}e^{- n^2at},\\
\fl&=&\frac{1}{2\pi}+\frac{1}{2\pi}\sum_{n=-\infty}^{-1}\left(e^{in(\phi-
\phi_0)}+e^{-in(\phi-\phi_0)}\right)e^{- n^2at},
\end{eqnarray}
\end{numparts}
thus eventually yielding
\begin{equation}
P(\phi,t|\phi_0)=\frac{1}{\pi}\left(\frac{1}{2}+\sum_{n=1}^{\infty}\cos
n(\phi-\phi_0)\exp{\left(-n^2\frac{D}{v_0^2}t\right)}\right).
\label{pdf_phi}
\end{equation}
This formula is used below in \ref{appc}.

\section{Derivation of the MSD of the Mikhailov-Meink{\"o}hn model}
\label{appc}

The APP model was introduced in in \cite{Mikhailov1997SelfmotionIP}.
The expression of the MSD is obtained from the identity $\mathbf{r}
(t)=\int_0^t\mathbf{v}(t')dt'$, knowing that $\mathbf{v}(t)=v_0\mathbf{e_v}$,
with $\mathbf{e_v}=(\cos\phi,\sin\phi)$. This leads to
\begin{eqnarray}
\langle\mathbf{r}^2(t)\rangle&=&\langle(v_0\int_0^{t}\mathbf{e_v}(t_1)dt_1)
(v_0\int_0^{t}\mathbf{e_v}(t_2)dt_2)\rangle,\\
&=&v_0^2\int_0^t\int_0^t\langle\cos\phi(t')\rangle dt_1dt_2.
\label{integ}
\end{eqnarray}
Here $\phi$ is the angle between the direction of motion of the active agent
at $t=t_1$ and at $t=t_2$, and $t'$ is the time difference, either $t_1-t_2$
or $t_2-t_1$. Indeed, the probability density of the Wiener process $\phi$
is only defined for positive times, therefore $t'>0$. This means that $t'=t_2
-t_1$ for $t_2>t_1$ and $t'=t_1-t_2$ for $t_1>t_2$. We rewrite the integral
as
\begin{equation}
\fl\langle\mathbf{r}^2(t)\rangle=v_0^2\int_0^t\int_0^{t_1}\langle\cos\phi(t_1-
t_2)\rangle dt_1dt_2+v_0^2\int_0^t\int_{t_1}^t\langle\cos\phi(t_2-t_1)
\rangle dt_1dt_2.
\end{equation}
Changing the order of integration in the second term and using the fact that
the cosine is an even function, we get
\begin{equation}
\langle\mathbf{r}^2(t)\rangle=2v_0^2\int_0^t\int_0^{t_1}\langle\cos\phi(t_1-
t_2)\rangle dt_1dt_2.
\label{integ2}
\end{equation}
To compute the quantity $\langle\cos{\phi(t')}\rangle$ we use the property
that
\begin{equation}
\langle\cos\phi(t')\rangle=\int_0^{2\pi}\cos\phi(t')P(\phi,t')d\phi,
\label{av_cos}
\end{equation}
with
\begin{equation}
P(\phi,t')=\int_{-\pi}^{\pi}P(\phi,t'|\phi_0)P(\phi_0)d\phi_0.
\label{init_cond}
\end{equation}
We further note that $\phi$ is a $2\pi$-periodic Wiener process, which is
distributed according to \eqref{pdf_phi}. Posing a delta distribution for
$\phi_0$, $P(\phi_0)=\delta(\phi_0)$ (it has to be precisely this
distribution as the angles are necessarily the same for $\phi(t'=0)=\phi
(t_1=t_2)$) we can then compute the ensemble average 
\begin{equation}
\fl\langle\cos\phi(t')\rangle=\frac{1}{\pi}\left[\frac{1}{2}\int_0^{2\pi}
\cos\phi d\phi+\int_0^{2\pi}\sum_{n=1}^{\infty}\cos\phi\cos n\phi\exp
\left(-n\frac{D}{v_0^2}t'\right)d\phi\right].
\end{equation}

The term on the left cancels and $\cos\phi\cos n\phi=\frac{1}{2}(\cos(n+1)
\phi+\cos(n-1)\phi)$, whose integral from 0 to $2\pi$ is always 0, except
for $n=1$, which then yields $\pi$. We thus find
\begin{equation}
\langle\cos\phi(t')\rangle=\exp\left(-\frac{D}{v_0^2}t'\right),
\end{equation}
which we may now plug into equation \eqref{integ2}. We calculate the
double integral,
\begin{eqnarray}
&&\int_0^t\int_0^{t_1}\langle\cos\phi(t_1-t_2)\rangle dt_1dt_2=
\int_0^t\int_0^{t_1}\exp\left(\frac{D}{v_0^2}(t_2-t_1)\right)dt_1dt_2,
\nonumber\\
&=&\int_0^t\left(\frac{v_0^2}{D}\right)\left(1-\exp\left(-\frac{D}{
v_0^2}t_1\right)\right)dt_1,
\nonumber\\
&=&\frac{v_0^2}{D}\left[\int_0^tdt_1-\int_0^t\exp\left(-\frac{D}{v_0^2}t_1
\right)dt_1\right],
\nonumber\\
&=&\frac{v_0^2}{D}t-\frac{v_0^2}{D}\left(\frac{-v_0^2}{D}\right)\left(\exp
\left(-\frac{D}{v_0^2}t_1\right)-1\right),
\nonumber\\
&=&\frac{v_0^2}{D}t+\frac{v_0^4}{D^2}\left(\exp\left(-\frac{D}{v_0^2}t
\right)-1\right).
\label{part1}
\end{eqnarray}
We then obtain the final expression
\begin{equation}
\langle\mathbf{r}^2(t)\rangle=\frac{2v_0^4}{D}t+\frac{2v_0^6}{D^2}\left(
\exp\left(-\frac{D}{v_0^2}t\right)-1\right).
\label{MSDu_moi}
\end{equation}
This result differs slightly from that reported in
\cite{Mikhailov1997SelfmotionIP} (in their notation $D=\sigma/m^2)$, which
reads
\begin{equation}
\langle\mathbf{r}^2(t)\rangle=\frac{2v_0^4m^2t}{\sigma}+\frac{v_0^6m^4}{
\sigma^2}\left(\exp\left(\frac{-2\sigma t}{m^2v_0^2}\right)-1\right).
\label{MSD_U1}
\end{equation}
Thus, equation \eqref{MSDu_moi} has an extra factor of 2 in front of the
$v_0^6$-term and a missing factor of 2 in the exponential, as compared to
results in \cite{Mikhailov1997SelfmotionIP}. However, both results have the
same long-time limit, $D_{\mathrm{eff}}=\frac{v_0^4}{2D}$. In contrast, our
corrected result has the short time limit $v_0^2t^2$ (instead of $2v_0^2t^2$).

\section{Applications of the Fox $H$-function technique and asymptotic
expansions}
\label{appfox}

We here provide details on the application of the Fox $H$-function technique
to calculate the integrals of the main part, for both the APP and ABP cases.

\subsection{Active particle with passive-noise case}
\label{appc1}

We start with equation 
\begin{equation}
\label{NumerInt}
P(\mathbf{r},t)=\frac{1}{4\pi t}\int_0^{\infty}\frac{1}{D_{\mathrm{
eff}}}p(D_{\mathrm{eff}})\exp\left(-\frac{\mathbf{r}^2}{4D_{\mathrm{
eff}}t}\right)dD_{\mathrm{eff}},
\end{equation}
for which we know that (where $k$, $\beta$, and $\gamma$ are the parameters
of the generalised Gamma distribution (\ref{pexp}))
\begin{equation}
p(D_{\mathrm{eff}};k,\theta,\beta)=\frac{\beta(2DD_{\mathrm{eff}})^{1/4}}{
4\theta\Gamma(k)D_{\mathrm{eff}}}\left(\frac{(2DD_{\mathrm{eff}})^{1/4}}{
\theta}\right)^{k\beta-1}e^{-((2DD_{\mathrm{eff}})^{1/4}/\theta)^\beta}.
\end{equation}
We introduce the abbreviations 
\begin{eqnarray}
a&=&\frac{(2D)^{\beta/4}}{\theta^{\beta}},
\label{a}\\
s&=&aD_{\mathrm{eff}}^{\beta/4},
\label{s}\\
l&=&\frac{4}{\beta},
\label{l}\\
b&=&\frac{a^l\mathbf{r}^2}{4t},
\label{b}
\end{eqnarray}
so that we can rewrite and rearrange \eqref{NumerInt} in the form
\begin{eqnarray}
P(\mathbf{r},t)&=&\frac{1}{4\pi t}\frac{1}{\Gamma(k)}a^{l-1}\int_0^{\infty}
s^{k-l-1}\exp{(bs^{-l})}\exp{(-s)}ds.
\label{pdf1}
\end{eqnarray}
We now use the property \cite{mathai} (p.151)
\begin{equation}
e^{-z}=H_{0,1}^{1,0}\left[z\left|\begin{array}{l}\rule{1.2cm}{0.01cm}\\
(0,1)\end{array}\right.\right],
\end{equation}
meaning that \eqref{pdf1} can be rewritten as 
\begin{equation}
P(\mathbf{r},t)=\frac{a^{l-1}}{4\pi t\Gamma(k)}\int_0^{\infty}s^{k-l-1}
H_{0,1}^{1,0}\left[bs^{-l}\left[\begin{array}{l}\rule{1.2cm}{0.01cm}\\
(0,1)\end{array}\right.\right]\exp{(-s)}ds.
\end{equation}

We continue by using the identity \cite{mathai} (p.4 equation (1.2.2)) 
\begin{equation}
H_{p,q}^{m,n}\left[z\left|\begin{array}{l}(a_p,A_p)\\(b_q,B_q)\end{array}
\right.\right]=H_{q,p}^{n,m}\left[\frac{1}{z}\left|\begin{array}{l}(1-b_q,
B_q)\\(1-a_p,A_p)\end{array}\right.\right],
\label{FH1}
\end{equation}
which in our case leads to
\begin{equation}
H_{0,1}^{1,0}\left[bs^{-l}\left|\begin{array}{l}\rule{1.2cm}{0.01cm}\\(0,1)
\end{array}\right.\right]=H_{1,0}^{0,1}\left[\frac{s^l}{b}\left|
\begin{array}{l}(1,1)\\\rule{1.2cm}{0.01cm}\end{array}\right.\right].
\end{equation}

It can also be shown that (see \cite{prudnikov} p.300)
\begin{eqnarray}
\nonumber
\int_0^{\infty}&x^{\alpha-1}e^{-\sigma x}H_{p,q}^{m,n}\left[wx^r\left|
\begin{array}{l}(a_p,A_p)\\(b_q,B_q)\end{array}\right.\right]dx\\
&=\sigma^{
-\alpha}H_{p+1,q}^{m,n+1}\left[\frac{w}{\sigma^r}\left|\begin{array}{l}
(1-\alpha,r),(a_p,A_p)\\(b_q,B_q)\end{array}\right.\right].
\label{foxh1}
\end{eqnarray}
Making the appropriate replacements $\alpha=k-l$, $w=1/b$, $r=l$ and $\sigma
=1$ we eventually obtain
\begin{eqnarray}
\nonumber
P(\mathbf{r},t)&=&\frac{a^{l-1}}{4\pi t\Gamma(k)}H_{2,0}^{0,2}\left[\frac{1}{
b}\left|\begin{array}{l}(1-k+l,l),(1,1)\\\rule{1.2cm}{0.01cm}\end{array}\right.
\right]\\
&=&\frac{a^{l-1}}{4\pi t\Gamma(k)}H_{0,2}^{2,0}\left[b\left|\begin{array}{l}
\rule{1.2cm}{0.01cm}\\(k-l,l),(0,1)\end{array}\right.\right].
\label{before_last1}
\end{eqnarray}

We write \eqref{before_last1} explicitly with the parameters given in
equations \eqref{a} to \eqref{b}, yielding
\begin{equation}
P(\mathbf{r},t)=\frac{1}{4\pi t\Gamma(k)}\frac{(2D)^{1-4/\beta}}{\theta^{4-
\beta}}H_{0,2}^{2,0}\left[\frac{2D\mathbf{r}^2}{\theta^44t}\left|
\begin{array}{l}\rule{1.2cm}{0.01cm}\\(k-4/\beta,4\beta),(0,1)\end{array}
\right.\right].
\label{before_last2}
\end{equation}
To get the marginal $P_X(x,t)$ we write
\begin{eqnarray}
\nonumber
P_X(x,t)&=&\int_{-\infty}^{+\infty}P(x,y,t)dy,\\
\nonumber
&=&\frac{1}{4\pi t}\int_{-\infty}^{\infty}\int_0^{\infty}\frac{1}{
D_{\mathrm{eff}}}p(D_{\mathrm{eff}})\exp({-(x^2+y^2)/4D_{\mathrm{eff}}t})
dD_{\mathrm{eff}}dy,\\
\nonumber
&=&\frac{1}{4\pi t}\int_0^{\infty}\frac{1}{D_{\mathrm{eff}}}p(D_{\mathrm{
eff}})\exp({-x^2/4D_{\mathrm{eff}}t})dD_{\mathrm{eff}}\\
&&\times\int_{-\infty}^{\infty}\exp({-y^2/4D_{\mathrm{eff}}t})dy,
\end{eqnarray}
which, with the Gaussian integral $\int_{-\infty}^{\infty}e^{-ax^2}dx=\sqrt{
\frac{\pi}{a}}$, leads to
\begin{equation}
P_x(x,t)= \frac{1}{\sqrt{4\pi t}}\int_0^{\infty}\frac{1}{\sqrt{D_{\mathrm{
eff}}}}p(D_{\mathrm{eff}})\exp({-x^2/4D_{\mathrm{eff}}t})dD_{\mathrm{eff}}.
\label{Px}
\end{equation}
We then establish that 
\begin{equation}
P_X(x,t)=\frac{a^{l-1}}{\sqrt{4\pi t}\Gamma(k)}\int_0^{\infty}s^{k-l-1/
2}\exp{(bs^{-l})}\exp{(-s)}ds,
\end{equation}
which, using the expressions \eqref{FH1} and \eqref{foxh1}, and with
$\alpha=k-l+1/2$ leads to 
\begin{eqnarray}
\nonumber
P_X(x,t)&=&\frac{1}{\sqrt{4\pi t}}\frac{1}{\Gamma(k)}a^{l-1}H_{2,0}^{0,2}
\left[\frac{1}{b}\left|\begin{array}{l}(\frac{1}{2}-k+l,l),(1,1)\\
\rule{1.2cm}{0.01cm}\end{array}\right.\right]\\
&=&\frac{1}{\sqrt{4\pi t}}
\frac{1}{\Gamma(k)}a^{l-1}H_{0,2}^{2,0}\left[b\left|\begin{array}{l}
\rule{1.2cm}{0.01cm}\\(k-l+\frac{1}{2},l),(0,1)\end{array}\right.\right].
\label{before_last_marg1}
\end{eqnarray}
An asymptotic expansion of the previous expression can be found in
\cite{mathai2} (p.20, equation (1.108)), which for our special case of the
$H$-function with $m=q$ and $n=0$ produces
\begin{equation}
H_{q,0}^{p,q}(z)=\mathcal{O}(z^{(\mathrm{Re}(\delta)+1/2)/\mu})\exp\left(
-\mu\beta^{-1/\mu}z^{1/\mu}\right),\quad |z|\to\infty,
\end{equation}
where $\delta$, $\mu$, and $\beta$ are parameters related to the parameters
$a_i$, $A_i$, $b_j$, $B_j$ of the Fox $H$-function (see \cite{mathai2} p.3).
In our case, we get, as a functions of the parameters defined above, 
\begin{eqnarray}
\beta&=&l^l,\\
\mu&=&1+l,\\
\delta&=&k-l-\frac{1}{2}.
\end{eqnarray}
Making the appropriate replacements, we may therefore get an asymptotic
expression valid for large displacements of the marginal displacement
PDF \eqref{Px}, namely,
\begin{eqnarray}
\nonumber
P_X(x,t)&\simeq&\frac{a^{l-1}}{\sqrt{4\pi t}\Gamma(k)}\left(\frac{a^lx^2}{4t}
\right)^{(k-l)/(1+l)}\\
&&\times\exp\left(-\frac{1+l}{l^{l/(1+l)}}\left(\frac{a^lx^2}{
4t}\right)^{1/(1+l)}\right),\quad\frac{a^lx^2}{4t}\to\infty.
\label{marg_fin1}
\end{eqnarray}

\subsection{Active Brownian particles}

For the ABP model we follow the same procedure as in the previous section.
We point out that we here take $D_{\mathrm{eff}}\simeq v^2/2D_R$, as the
translational diffusivity $D_T$ is neglected in this calculation, see the
main text. We now have that (with $k$, $\beta$, $\gamma$ as the parameters
of the generalised Gamma distribution (\ref{pexp}))
\begin{equation}
\fl p(D_{\mathrm{eff}};k,\theta,\beta)=\frac{\beta(2D_RD_{\mathrm{eff}})^{1/2}}{2\theta\Gamma(k)D_{\mathrm{eff}}}\left (\frac{(2D_RD_{\mathrm{eff}})^{1/
2}}{\theta}\right)^{k\beta-1}e^{-((2D_RD_{\mathrm{eff}})^{1/2}/\theta)^\beta}.
\end{equation}
We then introduce the abbreviations
\begin{eqnarray}
a&=&\frac{(2D_R)^{\beta/2}}{\theta^{\beta}},
\label{a1}\\
s&=&aD_{\mathrm{eff}}^{\beta/2},
\label{s1}\\
l&=&\frac{2}{\beta},
\label{l1}\\
b&=&\frac{a^l\mathbf{r}^2}{4t},
\label{b1}
\end{eqnarray}
The next steps are exactly the same as above in \ref{appc1}, such that we can
directly write
\begin{equation}
P(\mathbf{r},t)=\frac{1}{4\pi t\Gamma(k)}\frac{(2D_R)^{1-2/\beta}}{\theta^{2-
\beta}}H_{0,2}^{2,0}\left[\frac{2D_R\mathbf{r}^2}{\theta^24t}\left|
\begin{array}{l}\rule{1.2cm}{0.01cm}\\(k-2/\beta,2\beta),(0,1)\end{array}
\right.\right].
\label{before_lastr3}
\end{equation}

For the marginal, we use \eqref{marg_fin1} which for the distribution found
experimentally in \cite{beta}), where $l=2$ ($\beta=1$) and $k=2$, and find
\begin{equation}
\fl P_X(x,t)\simeq\frac{1}{\sqrt{4\pi t}}\frac{1}{\Gamma(2)}\frac{\sqrt{2D_R}}{
\theta}\exp\left(-3\times2^{-2/3}\left(\frac{\sqrt{2D_R}x^2}{4t\theta}\right)
^{1/3}\right),\quad\frac{D_Rx^2}{2t\theta^2}\to\infty.
\label{Px_ABP_paper1}
\end{equation}
For the case of a Rayleigh speed PDF, for which $l=1$ ($\beta=2$), $k=1$,
and $\theta=\sqrt{2D_RD_{\mathrm{eff}\star}}$, we find
\begin{equation}
P_X(x,t)\simeq\frac{1}{\sqrt{4\pi D_{\mathrm{eff}\star} t}\Gamma(2)}\exp
\left(-\left(\frac{x^2}{2D_{\mathrm{eff}\star}}\right)^{1/2}\right),
\quad\frac{x^2}{2D_{\mathrm{eff}\star}}\to\infty,
\label{Px_ABP_ray}
\end{equation}
or exactly the distribution that is obtained analytically with the reverse engineering procedure mentioned above.

\end{appendix}

\end{document}